\PassOptionsToPackage{dvipsnames}{xcolor}

\documentclass[submission, Phys]{SciPost}

\usepackage{amssymb}
\usepackage{bbm}
\usepackage{tikz}
\usetikzlibrary{calc,positioning,arrows.meta,fit,backgrounds}
\usepackage[ruled,vlined,linesnumbered]{algorithm2e}
\usepackage{microtype}

\hypersetup{colorlinks=true,linkcolor=Blue,citecolor=Blue,urlcolor=Blue}

\tikzset{
  wtensor/.style={draw, rounded corners=2pt, fill=Blue!12, minimum width=8mm, minimum height=6mm, inner sep=2pt, font=\small},
  vtensor/.style={draw, rounded corners=2pt, fill=Orange!25, minimum width=7mm, minimum height=6mm, inner sep=2pt, font=\small},
  ctensor/.style={draw, circle, fill=ForestGreen!35, minimum size=4.5mm, inner sep=0pt, font=\scriptsize},
  mpstensor/.style={draw, rounded corners=3pt, fill=ForestGreen!18, minimum width=12mm, minimum height=7mm, inner sep=1pt, font=\small},
  flatvec/.style={draw, fill=gray!20, minimum size=4.5mm, inner sep=1pt, font=\scriptsize},
  proj/.style={draw, fill=red!18, minimum size=4.5mm, inner sep=1pt, font=\scriptsize},
  leg/.style={thick},
  bond/.style={thick,gray!70!black},
}

\newcommand{\Wgate}{W}
\newcommand{\Vgate}{V}
\newcommand{\Ctensor}{C}
\newcommand{\onevec}{\mathbb{I}}

\begin{document}

\begin{center}{\Large \textbf{
Universal Information-Theoretic Structure of the Quasi-Stationary\\
Domany--Kinzel Automaton
}}\end{center}

\begin{center}
H.-Y. Lee\textsuperscript{1,2},
K. Harada\textsuperscript{3},
N. Kawashima\textsuperscript{2,4}
\end{center}

\begin{center}
{\bf 1} Department of Applied Physics, Graduate School, Korea University, Sejong 30019, Korea
\\
{\bf 2} Institute for Solid State Physics, The University of Tokyo, Kashiwa, Chiba 277-8581, Japan
\\
{\bf 3} Graduate School of Informatics, Kyoto University, Kyoto 606-8501, Japan
\\
{\bf 4} Trans-scale Quantum Science Institute, The University of Tokyo, Tokyo 113-0033, Japan
\end{center}

\begin{center}
\today
\end{center}

\section*{Abstract}
{\bf
We characterize the quasi-stationary distribution (QSD) of the bond directed-percolation line of the Domany--Kinzel automaton using a matrix-product-state representation of the probability distribution, obtained by projecting out the absorbing state and iterating the transfer matrix. Unlike moment- or sampling-based methods, this yields the full conditional distribution and direct access to information-theoretic diagnostics. The spatial structure of the QSD changes sharply across the transition: the active phase is bulk-like with finite density, whereas in the inactive phase the surviving activity collapses into a single flock --- the smallest interval containing all active sites --- occupying a vanishing fraction of the chain. Throughout the inactive phase the bipartite mutual information of the QSD equals the entropy of a single binary choice --- whether the flock lies to the left or right of the cut --- so the surviving clusters together encode just one bit of positional information, corresponding to a single effective cluster.
}

\vspace{10pt}
\noindent\rule{\textwidth}{1pt}
\tableofcontents\thispagestyle{fancy}
\noindent\rule{\textwidth}{1pt}
\vspace{10pt}

\section{Introduction}
\label{sec:intro}
Directed percolation (DP) is the paradigmatic universality class for continuous transitions into absorbing states~\cite{Janssen1981,Grassberger1982,Hinrichsen2000}.
The Domany--Kinzel (DK) automaton~\cite{Domany1984} provides a minimal one-dimensional realization of this physics.
It has long served as a benchmark for numerical and theoretical approaches to nonequilibrium critical phenomena.
On the bond-DP line, the model has an active phase with nonzero stationary density in the thermodynamic limit and an inactive phase in which activity eventually dies out.
For any finite chain, however, the unique stationary state of the DK dynamics is the all-inactive absorbing configuration, which carries no physical information. The object of interest is instead the quasi-stationary distribution (QSD) --- the distribution conditioned on non-absorption~\cite{Dickman2002,Oliveira2005}.
In the active phase, where extinction is exponentially rare in the system size, the QSD nearly coincides with the bulk stationary state of the original dynamics. In the inactive phase, where extinction occurs on a polynomial time scale, the conditioning on survival becomes a strong selection and the QSD acquires a qualitatively distinct, fluctuation-stabilized structure.

Most numerical studies of absorbing-state systems characterize the QSD through low-order observables such as the density, lifetime, and moment ratios~\cite{Hinrichsen2000,Dickman2002,Oliveira2005}.
Here we ask how typical surviving configurations are organized at the level of the full joint distribution: independent active sites, compact clusters, and a collectively delocalized cluster can share the same vanishing mean density yet represent distinct spatial organizations that low-order observables alone cannot resolve.
The relevant object is therefore the spatial structure of the conditional measure itself.

Conventional methods are not well suited to these distribution-level questions. Monte Carlo sampling is highly effective for typical observables, but it yields configurations drawn from the distribution rather than the distribution itself. Rare configurations are therefore sampled exponentially rarely, if at all, so their probabilities are not captured—the absorbing state in the active phase and strongly active configurations in the inactive phase remain effectively out of reach. More fundamentally, without explicit access to the configuration probabilities, entropic measures such as the Shannon and Rényi entropies of the QSD are not directly accessible from samples alone—all the more so when they are dominated by rare configurations—since they are built from the probabilities themselves rather than from sampled averages.

To overcome this limitation and address the distribution-level question above, we develop a matrix-product-state (MPS) framework that obtains the QSD extending the algorithm proposed in Ref.~\cite{Harada2019}; this provides a compact representation of the full conditional distribution on chains and exact sample generation by sequential conditional sampling. The essential difference from Monte Carlo is that the MPS represents the probability $P_{\rm QS}(X)$ of each configuration explicitly, not merely a stream of samples; this direct access to configuration probabilities is what makes information-theoretic quantities such as entropies measurable, and it underlies the analysis that follows.
In particular, it provides a distribution-level characterization of the surviving activity in both the active and inactive phases. It also gives direct access to bipartite marginal entropies and hence to information-theoretic observables such as the bipartite Shannon mutual information of the QSD, allowing the surviving activity to be analyzed in information-theoretic terms. Such a diagnostic, recently introduced for unconditioned classical many-body dynamics~\cite{Pizzi2022,Pizzi2024}, has not, to our knowledge, been applied to the quasi-stationary (survival-conditioned) dynamics of absorbing-state models.

An information-theoretic analysis of the quasi-stationary distribution reveals that, in the inactive phase, the surviving activity is a single flock whose clusters share one collective position, so the bipartite mutual information reduces to a single bit of left--right positional information --- one effective cluster, even though a typical surviving configuration contains several distinct clusters. The active phase, by contrast, is a bulk-like state with no extended positional degree of freedom.

\begin{figure}[!t]
    \centering
    \includegraphics[width=0.8\linewidth]{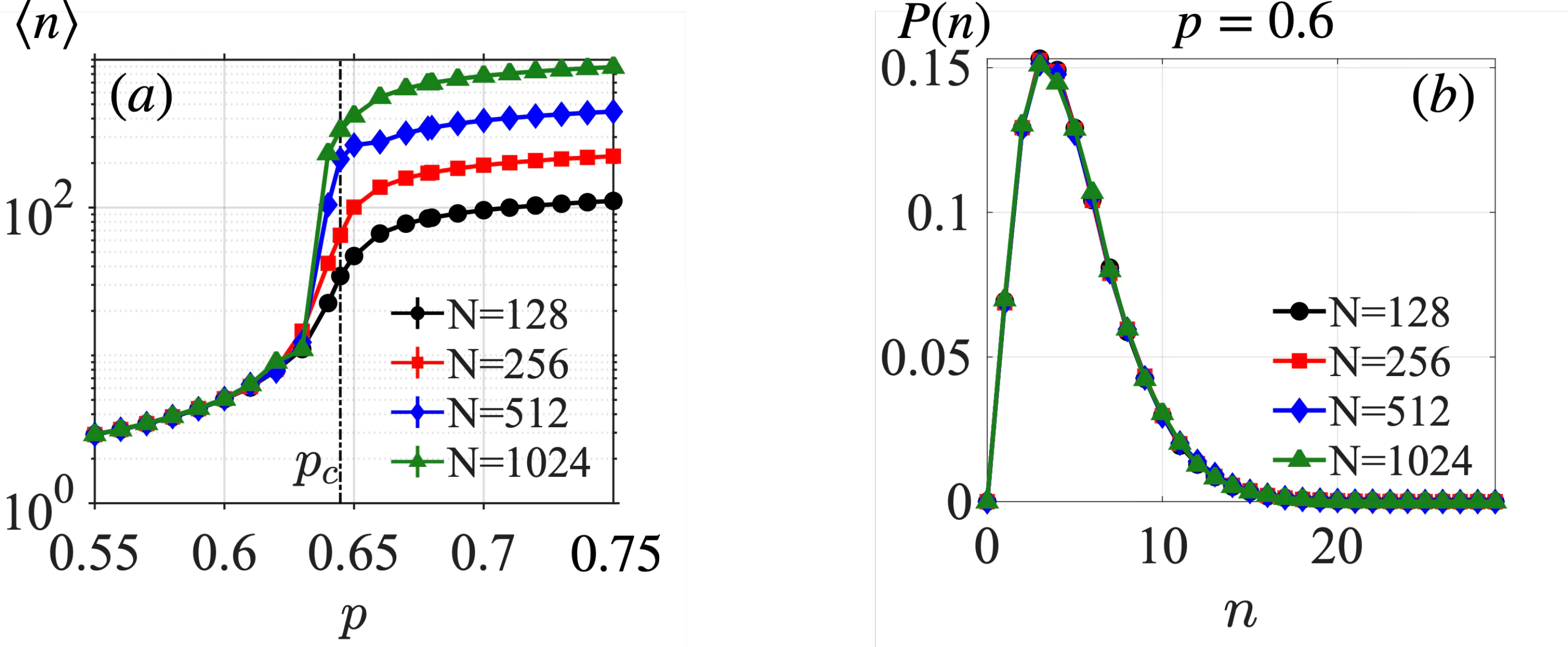}
    \caption{QSD statistics of the total active-site count $n=\sum_i x_i$.
    (a) mean active count $\langle n\rangle$ versus $p$ on a log scale, for $N=128,256,512,1024$.
    Across the inactive phase ($p<p_c$) the four curves collapse onto a common, $N$-independent value; across the active phase ($p>p_c$) the curves separate and $\langle n\rangle$ grows with $N$, consistent with a finite intensive density.
    A dashed vertical line marks $p_c$.
    (b) probability distribution $P(n)$ at $p=0.60$ for the same $N$ values; the curves collapse onto a common limit shape, showing that the $N$-independence of the active count holds at the level of the full distribution, not just the mean.}
    \label{fig:active-number}
\end{figure}

\section{Model and quasi-stationary distribution as a matrix-product eigenvector}
\label{sec:model}\label{sec:method}
The DK automaton evolves binary variables $x_i(t)\in\{0,1\}$ on a tilted square lattice of $N$ sites.
The probability that a site is active at the next time step, $P[n]$, depends on the number $n$ of active parents, $P[0]=0,\qquad P[1]=p_1,\qquad P[2]=p_2$.
We focus on the bond-DP line, $p_1=p$ and $p_2=p(2-p)$, whose critical point is $p_c=0.644700185(5)$~\cite{Jensen2004}.
The condition $P[0]=0$ makes $|0^N\rangle$ absorbing.

\subsection{Doob conditioning and the projected transfer matrix}
\label{sec:doob-cond}
Let $T$ be the stochastic transfer matrix for one DK update.
The projector $\Pi = \mathbbm{1} - |0^N\rangle\langle 0^N|$ removes the absorbing configuration $|0^N\rangle$. The QSD $|P_{\rm QS}\rangle$ is the leading non-negative right eigenvector of $\Pi T\Pi$ with eigenvalue $\lambda_1<1$, equivalently the Doob-conditioned stationary distribution~\cite{ChetriteTouchette2015,Collet2013}; $\lambda_1$ is the survival probability per DK step.
In what follows, all expectations $\langle\cdot\rangle_{\rm QS}$ and probabilities $P_{\rm QS}(\cdot)$ refer to this Doob-conditioned stationary distribution.

The projected operator $\Pi T\Pi$ thus targets the survival-conditioned stationary state directly. This is complementary to earlier matrix-product treatments of classical stochastic lattice dynamics, which construct exact steady states~\cite{Derrida1993}, evolve the full distribution in real time~\cite{Johnson2010,Harada2019}, or extract large-deviation statistics of time-integrated observables from the dominant eigenvector of a tilted generator~\cite{Banuls2019,Helms2019}: there the object of interest is a biased trajectory ensemble, whereas the projection selects the quasi-stationary regime of the absorbing dynamics itself.

\subsection{The probability MPS and the two-layer transfer matrix}
\label{sec:prob-mps}
The probability distribution is represented directly as an MPS,
    $ |P\rangle = \sum_X P(X)|X\rangle $,
so physical averages are computed with the flat left vector $\langle\mathbf{1}|=\sum_X\langle X|$, not with the Born-rule norm. The projection enforces $P_{\rm QS}(0^N)=0$, so $|P_{\rm QS}\rangle$ is supported on the non-absorbing configurations and is normalized by $\langle\mathbf{1}|P_{\rm QS}\rangle=\sum_X P_{\rm QS}(X)=1$; the absorbing state contributes zero, so every configuration sum $\sum_X$ below effectively runs over the non-absorbing configurations.

One DK time step factorizes into two brick-wall sub-steps generated by a three-leg bulk kernel $\Wgate$, a two-leg boundary kernel $\Vgate$, and copy tensors $\Ctensor$; Fig.~\ref{fig:S-2step} shows the resulting two-layer network, whose explicit construction is given in Appendix~\ref{sec:S1}. The transfer matrix is applied gate by gate with an immediate SVD recompression after each gate, using $\chi_{\max}=120$ and $\varepsilon_{\rm cut}=10^{-15}$ throughout (Appendix~\ref{sec:S2}); deep in the inactive phase the post-truncation bond dimension saturates well below $\chi_{\max}$ on chains up to $N=1024$, so the projected iteration remains inexpensive in the regime of interest.

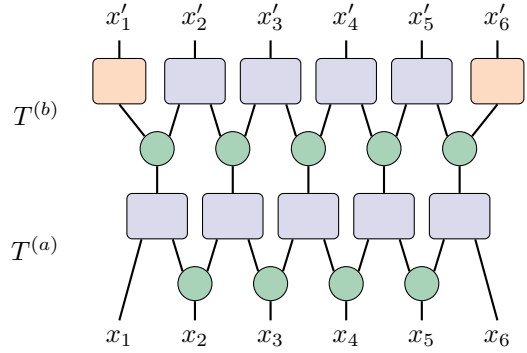
\begin{figure}[h]
\centering
\begin{tikzpicture}[x=1.0cm,y=0.85cm]
\foreach \i in {1,...,6} {\node[font=\small] (x\i) at (\i,-1.0) {$x_{\i}$};}
\foreach \i in {2,3,4,5} {\node[ctensor] (C\i) at (\i,-0.15) {};}
\foreach \i in {1,...,5} {%
  \pgfmathsetmacro{\xpos}{\i+0.5}
  \node[wtensor] (W\i) at (\xpos,0.9) {};
}
\foreach \i in {1,...,5} {%
  \pgfmathsetmacro{\xpos}{\i+0.5}
  \node[ctensor] (D\i) at (\xpos,1.95) {};
}
\node[vtensor] (Vl) at (1,3.0) {};
\node[vtensor] (Vr) at (6,3.0) {};
\foreach \i in {1,...,4} {%
  \pgfmathsetmacro{\xpos}{\i+1}
  \node[wtensor] (Wb\i) at (\xpos,3.0) {};
}
\foreach \i in {1,...,6} {\node[font=\small] (xp\i) at (\i,4.0) {$x'_{\i}$};}
\foreach \i in {2,3,4,5} {\draw[leg] (x\i) -- (C\i.south);}
\draw[leg] (x1.north) -- ([xshift=2mm]W1.south west);
\draw[leg] (x6.north) -- ([xshift=-2mm]W5.south east);
\draw[leg] (C2.north west) -- ([xshift=-2mm]W1.south east);
\draw[leg] (C2.north east) -- ([xshift=2mm]W2.south west);
\draw[leg] (C3.north west) -- ([xshift=-2mm]W2.south east);
\draw[leg] (C3.north east) -- ([xshift=2mm]W3.south west);
\draw[leg] (C4.north west) -- ([xshift=-2mm]W3.south east);
\draw[leg] (C4.north east) -- ([xshift=2mm]W4.south west);
\draw[leg] (C5.north west) -- ([xshift=-2mm]W4.south east);
\draw[leg] (C5.north east) -- ([xshift=2mm]W5.south west);
\foreach \i in {1,...,5} {\draw[leg] (W\i.north) -- (D\i.south);}
\draw[leg] (D1.north west) -- (Vl.south);
\draw[leg] (D1.north east) -- ([xshift=2mm]Wb1.south west);
\draw[leg] (D2.north west) -- ([xshift=-2mm]Wb1.south east);
\draw[leg] (D2.north east) -- ([xshift=2mm]Wb2.south west);
\draw[leg] (D3.north west) -- ([xshift=-2mm]Wb2.south east);
\draw[leg] (D3.north east) -- ([xshift=2mm]Wb3.south west);
\draw[leg] (D4.north west) -- ([xshift=-2mm]Wb3.south east);
\draw[leg] (D4.north east) -- ([xshift=2mm]Wb4.south west);
\draw[leg] (D5.north west) -- ([xshift=-2mm]Wb4.south east);
\draw[leg] (D5.north east) -- (Vr.south);
\draw[leg] (Vl.north) -- (xp1);
\draw[leg] (Vr.north) -- (xp6);
\foreach \i in {1,...,4} {%
  \pgfmathtruncatemacro{\out}{\i+1}
  \draw[leg] (Wb\i.north) -- (xp\out);
}
\node[font=\small,anchor=east] at (0.35,0.4) {$T^{(a)}$};
\node[font=\small,anchor=east] at (0.35,2.5) {$T^{(b)}$};
\end{tikzpicture}
\caption{Canonical tensor-network representation of one DK time step on $N=6$ sites [Eq.~\eqref{eq:S-Tfull}], in the convention of Ref.~\cite{Harada2019}. Bottom row: input sites $x_i$. Sub-step $a$ ($T^{(a)}$): $\Ctensor$ tensors duplicate internal inputs into the brick wall of bulk $\Wgate$ gates that produces the intermediates $m_k$ (internal bonds, not labelled). Sub-step $b$ ($T^{(b)}$): a second row of $\Ctensor$ tensors duplicates the intermediates into the layer-2 operators---boundary $\Vgate$ at the two ends, bulk $\Wgate$ in between---which generate the output sites $x'_i$.}
\label{fig:S-2step}
\end{figure}

\subsection{Power iteration and convergence}
\label{sec:power-iter}
We obtain $|P_{\rm QS}\rangle$ by power iteration on $T$, removing the absorbing component after each step and renormalizing.
Convergence was monitored by two criteria: (i) the eigenvector overlap satisfied $1-|\langle P_t|P_{t-\Delta t}\rangle|<10^{-6}$ after two-norm normalization, and (ii) the eigenvalue $\lambda_1$ varied by less than $10^{-4}$ between successive iterations, providing an additional stability check in the inactive phase. The explicit transfer-matrix construction, projection procedure, perfect-sampling algorithm, converged $\lambda_1$ values, and further numerical details (including bond-dimension choices) are provided in Appendices~\ref{sec:S1}--\ref{sec:S6}.

\subsection{Perfect sampling and distribution-level estimators}
\label{sec:perfect-sampling}
A practical advantage of obtaining the QSD as an MPS is that it admits exact, sequential perfect sampling~\cite{Harada2025}: configurations are drawn site-by-site from the QSD without Monte Carlo thermalization or rejection, giving direct access to distribution-level diagnostics on chains where brute-force simulation would be costly. 

With the right environments cached once per MPS, drawing a sample---and, as used below, evaluating its exact probability---costs $O(N\chi^2)$ per sample (Appendix~\ref{sec:S4}). Sequential sampling supplies more than configurations. Because each site is drawn from its exact conditional, the chain rule yields the exact probability of every sample, and the same site-by-site conditionals give $\log_2 P_{\rm QS}(x_{R_\ell}\mid x_{L_\ell})$ at every cut $\ell$ of a sample, where $L_\ell=\{1,\ldots,\ell\}$ and $R_\ell=\{\ell{+}1,\ldots,N\}$. The bipartite mutual information analysed below [Eq.~\eqref{eq:mi-def}] is therefore evaluated as an ordinary sample average,
\begin{equation}
I(L_\ell\!:\!R_\ell) \;=\; \Bigl\langle\, \log_2 P_{\rm QS}(x_{R_\ell}\mid x_{L_\ell}) \;-\; \log_2 P_{\rm QS}(x_{R_\ell}) \,\Bigr\rangle_{\rm QS},
\label{eq:mi-estimator}
\end{equation}
where the marginal $P_{\rm QS}(x_{R_\ell})$ is obtained for each sample by one exact MPS contraction against the flat covector on $L_\ell$. Both logarithms are exact for every sample, so the estimator is unbiased and its statistical error follows from the sample variance.

This per-sample likelihood is the step with no counterpart in sampling-only approaches. A plug-in estimate of the marginal entropies in Eq.~\eqref{eq:mi-def} would require the $2^{\ell}$ probabilities of an $\ell$-site marginal, out of reach at $\ell\sim N/2$ for any sample budget; with exact configuration probabilities available, each entropy instead reduces to a simple average of $-\log_2 P$, free of plug-in bias. Quasi-stationary simulation schemes~\cite{Dickman2002,Oliveira2005} generate configurations distributed according to the QSD but not the probabilities of those configurations, so the information-theoretic diagnostics used in this work are structurally out of their reach.

All data below use open chains of length $N=128,256,512,1024$.
We compute the QSD at $p=0.60, 0.61, \ldots, 0.72$, together with the critical point, drawing $10^5$ exact samples at each.

\section{Active-count statistics and local clustering}
\label{sec:active-count}\label{sec:clustering}
We characterize the QSD through the distribution of the total active-site count
\begin{align}
    n=\sum_i x_i,
    \qquad
    P(n)=\sum_X P_{\rm QS}(X)\,\delta_{n,\sum_i x_i}.
    \nonumber
\end{align}
Figure~\ref{fig:active-number} (left) shows the mean active count $\langle n\rangle=\sum_n n\,P(n)$ as a function of $p$ for $N=128,256,512,1024$.
Across the inactive phase the four curves collapse onto a common, $N$-independent value: the surviving activity contains $O(1)$ active sites regardless of system size.
The full distribution $P(n)$, shown in Fig.~\ref{fig:active-number} (right) at $p=0.60$, collapses onto a single limit shape for all $N$ studied --- showing that the $N$-independence of the active count holds at the level of the full distribution, not just the mean. This inactive saturation raises an immediate question: how is this fixed $O(1)$ total active mass spatially distributed within an increasingly long chain?
By contrast, in the active phase, $\langle n\rangle$ grows with $N$, recovering the conventional bulk picture in which the density $\bar\rho=\langle n\rangle/N$ approaches a finite intensive value.

\begin{figure}[!t]
    \centering
    \includegraphics[width=0.8\linewidth]{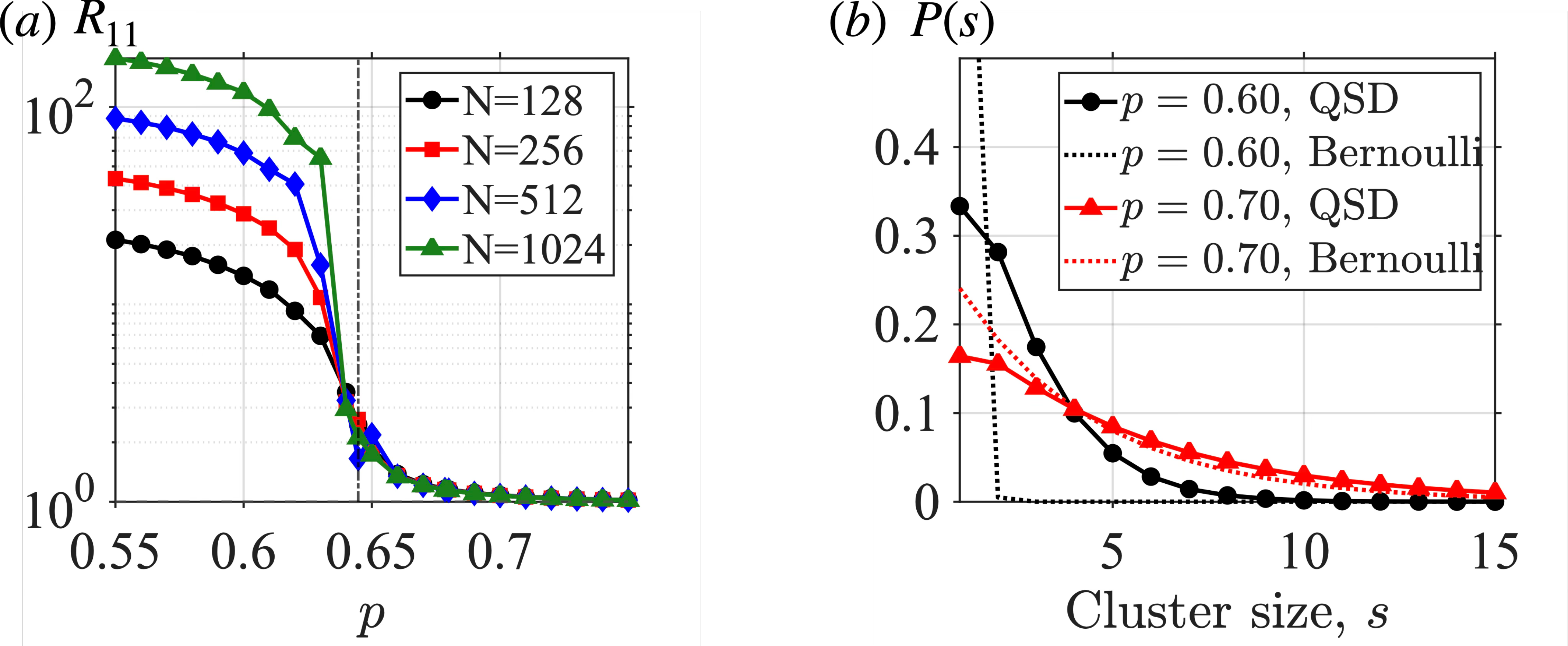}
    \caption{Clustering signatures of the QSD.
    (a) Nearest-neighbor clustering ratio $R_{11}$ [Eq.~\eqref{eq:r11-def}] versus $p$ for $N=128, 256, 512, 1024$.
    In the inactive phase $R_{11}\gg 1$ and grows with $N$; in the active phase $R_{11}\to 1$, the value expected for independent active sites.
    The vertical dashed line marks $p_c$.
    (b) Cluster-size distribution $P(s)$ at $p=0.60$ (inactive, black) and $p=0.70$ (active, red), each compared with the Bernoulli baseline $(1-\bar\rho)\bar\rho^{s-1}$ of the same one-site density (dotted).
    The inactive distribution differs sharply from Bernoulli, signaling strong clustering, while the active distribution nearly coincides with Bernoulli, consistent with a weakly correlated bulk gas.}
    \label{fig:p11-ratio}
\end{figure}

Beyond the size-scaling of the active count established above, density alone does not specify how the surviving activity is spatially organized within the chain.
To distinguish a gas of independent active sites from clustered activity, we measure the nearest-neighbor active-pair density relative to an independent-site (Bernoulli) baseline with the same one-site density profile,
\begin{align}
    R_{11} = \frac{\sum_{i=1}^{N-1}\langle x_i x_{i+1}\rangle_{\rm QS}}
                  {\sum_{i=1}^{N-1}\langle x_i\rangle_{\rm QS}\langle x_{i+1}\rangle_{\rm QS}}.
    \label{eq:r11-def}
\end{align}
Figure~\ref{fig:p11-ratio}(a) shows that $R_{11}$ is sharply phase-specific: in the active phase $R_{11}\approx 1$, the value expected for independent sites at the same density, while in the inactive phase $R_{11}\gg 1$ and grows approximately linearly with $N$\footnote{This $N$-scaling reflects the vanishing density baseline $\sum_i\langle x_i\rangle\langle x_{i+1}\rangle\sim N\bar\rho^2\sim 1/N$ in the inactive phase, with $\sum_i\langle x_i x_{i+1}\rangle$ remaining $O(1)$ by the $N$-independent saturation of $P(s)$ established below; it is not a signal of long-range correlation.} --- neighboring active sites are far from statistically independent. To translate this excess nearest-neighbor correlation into a direct geometric statement about how the surviving activity is organized, we measure the cluster structure of typical samples.

Each sample decomposes uniquely into clusters --- consecutive runs of active sites bounded on both sides by inactive sites or the chain endpoints --- and we let $P(s) = \langle n_s\rangle_{\rm QS}/\langle K\rangle_{\rm QS}$ denote the cluster-size distribution under the QSD, where $n_s(X)$ is the number of size-$s$ clusters in configuration $X$ and $K(X)=\sum_s n_s(X)$ is the total cluster count. Figure~\ref{fig:p11-ratio}(b) compares $P(s)$ at a representative inactive ($p=0.60$) and active ($p=0.70$) point with the same-density Bernoulli baseline $(1-\bar\rho)\bar\rho^{s-1}$ expected for independent active sites. In the active phase the two curves nearly coincide: the cluster-size statistics are accounted for by the one-site density alone. In the inactive phase $P(s)$ departs sharply from Bernoulli, signaling cluster correlations beyond what the density implies. At fixed inactive $p$, $P(s)$ is essentially $N$-independent across the chain sizes studied --- the cluster-size distribution settles to a limit shape with finite, $N$-independent support. These diagnostics indicate that the inactive QSD is not a dilute gas of independent active sites: the surviving activity organizes into finite-size clusters whose total mass remains $O(1)$ as the chain grows.

\section{Collective positional information}
\label{sec:positional}
Local clustering still leaves open whether the $O(1)$ surviving clusters are spread across the system or confined to a common region. The diagnostics below distinguish two pictures --- independently positioned clusters, whose left--right positions are uncorrelated, versus a single flock whose clusters are strongly correlated into one bound object sharing a single position (illustrated schematically in Appendix~\ref{sec:F-shape}).
We call the smallest interval containing all active sites the \emph{flock} of the configuration; its spatial extent
\begin{align}
    \ell_{\rm act}=\max\{i:x_i=1\}-\min\{i:x_i=1\}+1
\end{align}
characterizes how spread out the surviving activity is, while the intra-flock filling fraction
$
    \rho_{\rm flock}\equiv n_{\rm active}/\ell_{\rm act}
$
measures how densely the active sites pack within that interval.

\begin{figure}[!t]
    \centering
    \includegraphics[width=0.8\linewidth]{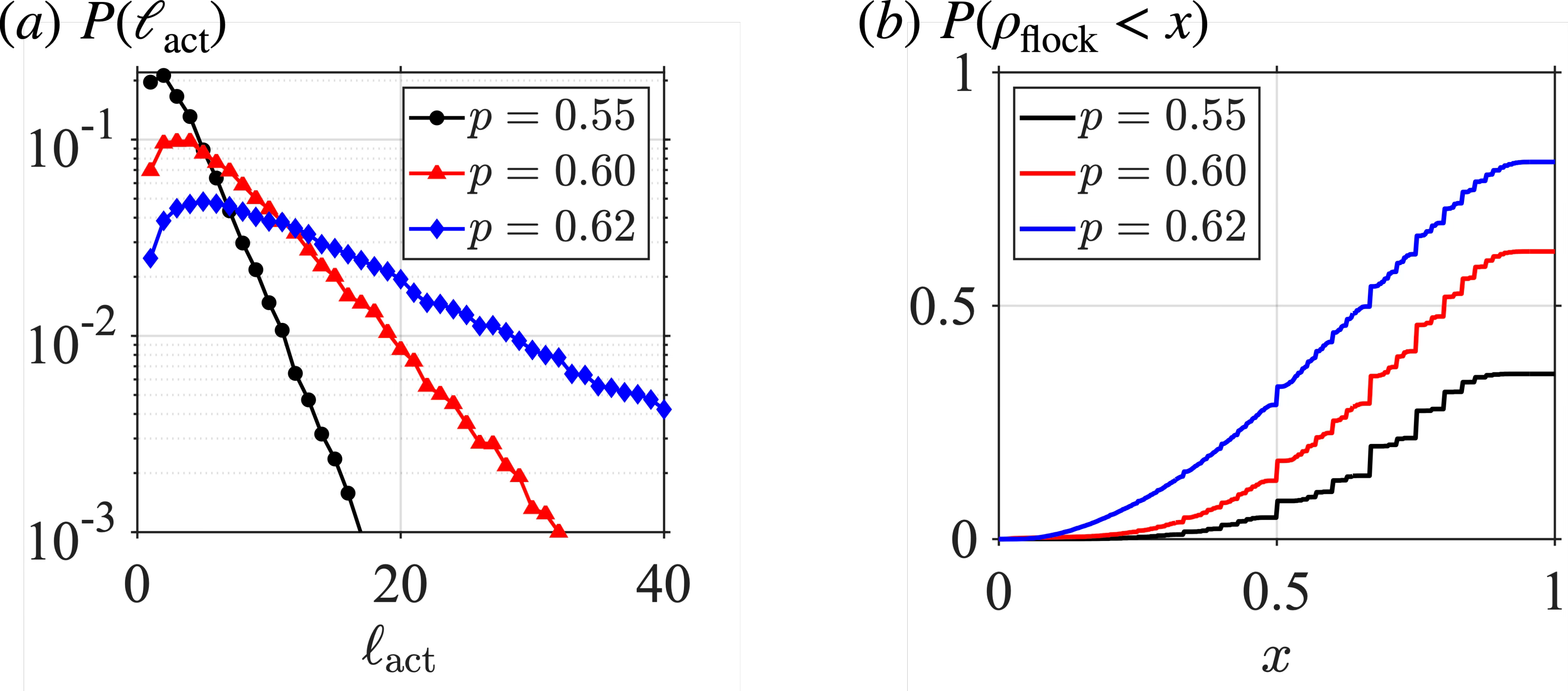}
    \caption{Flock morphology in the inactive phase at $p=0.55,~0.60,~0.62$ for $N=1024$.
    (a) Spatial-extent distribution $P(\ell_{\rm act})$: unimodal with a sharp peak at small $\ell$ that broadens and shifts to larger $\ell$ as $p\to p_c$, with an approximately exponential tail.
    (b) Cumulative filling distribution $P(\rho_{\rm flock}<x)$; the strict inequality excludes single-cluster samples, so the deficit from $1$ at $x=1$ equals the single-cluster fraction $P(\rho_{\rm flock}=1)$. Approaching $p_c$ this deficit shrinks ($0.64\to0.38\to0.19$) and the curve shifts to smaller $\rho_{\rm flock}$, i.e.\ the flock loosens.}
    \label{fig:l-rho-dist}
\end{figure}

The extent distribution $P(\ell_{\rm act})$ [Fig.~\ref{fig:l-rho-dist}\,(a)] is unimodal, with a peak at small $\ell_{\rm act}$ and an approximately exponential tail.
The peak broadens and shifts to larger $\ell_{\rm act}$ as $p$ approaches the critical value, but the typical extent stays much smaller than the chain length throughout.
The exponential tail indicates a finite correlation between the leftmost and rightmost surviving sites, consistent with a single bound object; independently distributed active regions would instead produce a broad, system-scale span distribution.

The filling is shown in Fig.~\ref{fig:l-rho-dist}\,(b) as the cumulative distribution $P(\rho_{\rm flock}<x)$; the strict inequality excludes the single-cluster configurations, so the offset of each curve from unity at $x=1$ equals $P(\rho_{\rm flock}=1)$, the probability that the flock is a single cluster. This probability is large deep in the inactive phase ($\simeq0.64$ at $p=0.55$), where the filling concentrates near unity, and it decreases monotonically toward $p_c$ as the distribution shifts to smaller $\rho_{\rm flock}$. The flock thus evolves from a single cluster deep in the inactive phase to an extended, partially filled sequence of clusters separated by holes near the transition, mirroring the divergence of the transverse correlation length $\xi_\perp(p)$, which enlarges the flock extent more rapidly than its active mass and itself follows the directed-percolation scaling $\xi_\perp\propto(p_c-p)^{-\nu_\perp}$ ($\nu_\perp\simeq1.097$~\cite{Jensen2004}; Appendix~\ref{sec:xi-perp}). Throughout, the flock remains a single connected object occupying a small fraction of the chain rather than a set of independently positioned clusters. 
This interpretation identifies the surviving activity with a single positional degree of freedom and gives the QSD a specific information-theoretic implication: treating the flock as one uniformly delocalized object should imprint a specific form on the bipartite statistics, which the MPS representation of QSD lets us investigate directly.

Concretely, we consider the bipartite Shannon mutual information of the QSD across a cut of the chain. A cut after site $\ell$ divides the chain into a left subsystem $L_\ell = \{1,\ldots,\ell\}$ and a right subsystem $R_\ell = \{\ell{+}1,\ldots,N\}$, and the bond-resolved value is
\begin{align}
    I(\ell,N) \equiv I(L_\ell\!:\!R_\ell)=H(L_\ell)+H(R_\ell)-H(L_\ell,R_\ell),
    \label{eq:mi-def}
\end{align}
where $H(L_\ell)=-\sum_{x_{L_\ell}}p(x_{L_\ell})\log_2 p(x_{L_\ell})$ is the Shannon entropy (in bits; all logarithms are base 2) of the marginal $p(x_{L_\ell})=\sum_{x_{R_\ell}}P_{\rm QS}(x_{L_\ell},x_{R_\ell})$ over $L_\ell$, with $H(R_\ell)$ and $H(L_\ell,R_\ell)$ defined analogously.
Assuming the flock is uniformly distributed over the chain, we derive the closed form
\begin{align}
    I(\ell, N) \simeq h\left(\ell/N\right) + \frac{K_{\mathrm{eff}}(p)}{N}\,\log_2\!\bigl[(\ell/N)(1-\ell/N)\bigr],
    \label{eq:mi-scaling}
\end{align}
throughout the inactive phase. Its leading term $h(\ell/N)=-(\ell/N)\log_2(\ell/N)-(1-\ell/N)\log_2(1-\ell/N)$ is the entropy of a single binary trial---which side of the cut contains the cluster---while the $O(1/N)$ correction has a universal Stirling-like shape $\log_2[x(1-x)]$, with $x\equiv\ell/N$, and a non-universal amplitude $K_{\mathrm{eff}}(p)$ set by the effective flock width. The derivation and a numerical check of both universal scaling functions and $K_{\mathrm{eff}}(p)$ are given in Appendix~\ref{sec:F-shape}.

\begin{figure}[!t]
    \centering
    \includegraphics[width=0.7\linewidth]{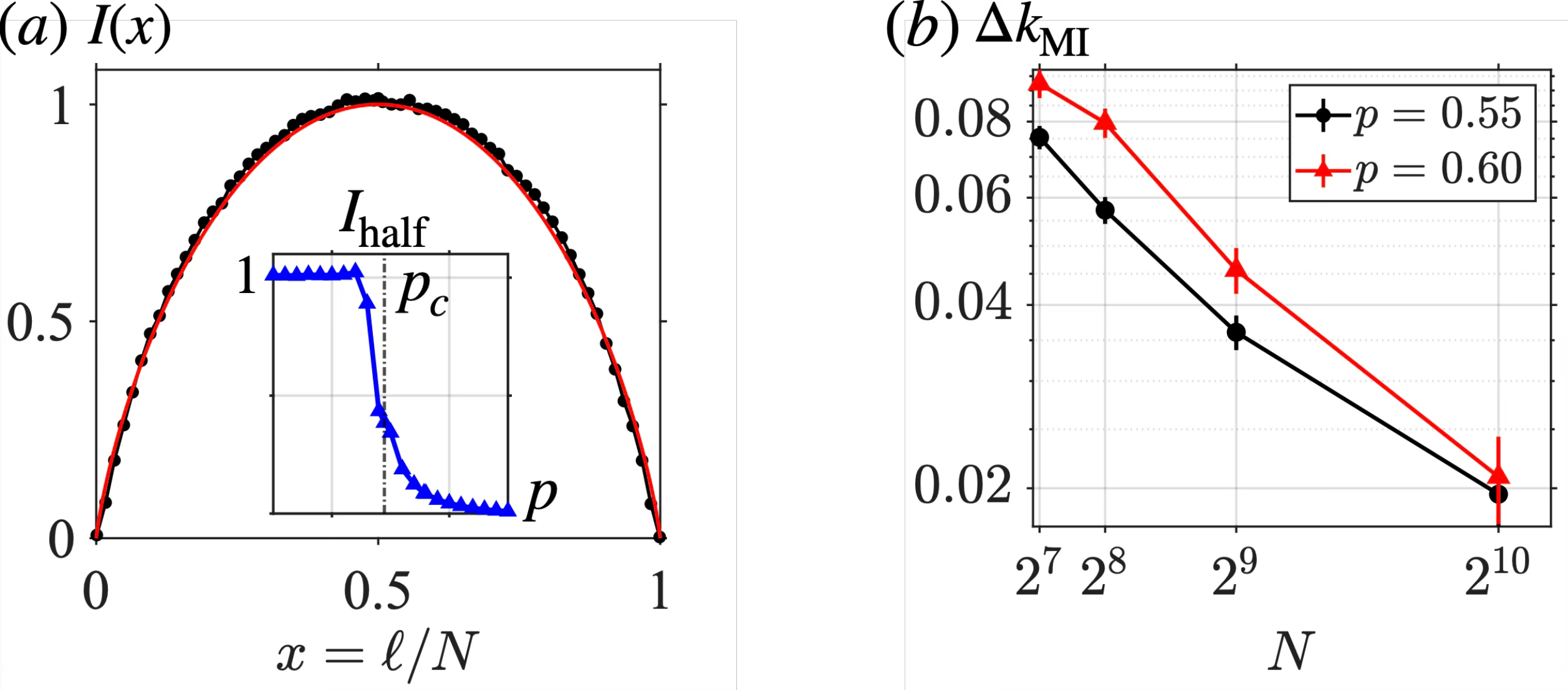}
    \caption{Mutual-information diagnostics of the QSD, with all entropies in bits ($\log_2$).
    (a) Bond-resolved $I(\ell,N)$ at $p=0.60$ and $N=1024$ (inactive) versus cut position $x=\ell/N$, closely following $h(x)$ (red) and peaking at $h(1/2)=1$.
    Inset: half-chain $I_{\rm half}\equiv I(N/2,N)$ versus $p$ at $N=1024$, with a plateau near $1$ in the inactive phase that drops sharply across $p_c$ (dash-dotted line) and decays in the active phase.
    (b) Finite-size scaling of $\Delta k_{\rm MI}\equiv k_{\rm MI}-1$ versus $N$ at $p=0.55$ and $0.60$, consistent with $k_{\rm MI}\to 1$ as $N\to\infty$.}
    \label{fig:mutual-info}
\end{figure}

Evaluating Eq.~\eqref{eq:mi-def} directly on $10^5$ perfect samples drawn from the QSD reproduces Eq.~\eqref{eq:mi-scaling} across the inactive phase.
Figure~\ref{fig:mutual-info}(a) shows $I(\ell, N)$ at $p=0.60$ and $N=1024$: the bond-resolved profile closely follows $h(\ell/N)$, the entropy of a single binary position, with the small residual deviation set by the $K_{\mathrm{eff}}(p)\log_2[x(1-x)]/N$ correction.
The active phase is qualitatively different: at $p=0.70$, $I(\ell, N)$ is essentially zero across all cuts (Appendix~\ref{sec:F-shape}), signalling that no extended positional degree of freedom survives.
The inset of Fig.~\ref{fig:mutual-info}(a) traces the half-chain value $I(N/2, N)$ at $N=1024$ across the transition: a plateau near $1 = h(1/2)$ runs through the inactive phase, drops sharply across $p_c$, and decays to small values deep in the active phase; its approach to exactly one bit as $N\to\infty$ is quantified in Fig.~\ref{fig:mutual-info}(b).

This agreement is notable given the flock morphology of Fig.~\ref{fig:l-rho-dist}: near $p_c$, extended flocks with low internal filling --- multiple clusters threaded by genuine internal holes --- carry non-negligible weight, yet the bipartite mutual information across all cuts behaves as if drawn from a single uniformly delocalized object.
The geometric multiplicity of contiguous active runs produces no additional positional degrees of freedom: the clusters share one collective position, that of the enclosing flock.

To make this single flock behavior quantitative, we ask how many effective independent positional degrees of freedom the MI actually carries, by comparing it to a reference model of $k$ clusters placed independently and uniformly over the chain.
Each cluster falls in $L_{N/2}$ with probability $1/2$, so the left-half cluster count is $n_L\sim\mathrm{Bin}(k,1/2)$.
The only shared information between halves is the constraint $n_L+n_R=k$, giving the calibration
\begin{align}
    I_k = H[\mathrm{Bin}(k,1/2)],
    \label{eq:bin-mi}
\end{align}
where $H[\mathrm{Bin}(k,1/2)]$ denotes the Shannon entropy of the binomial distribution with $k$ trials and success probability $1/2$.
$I_k$ increases monotonically in $k$ with $I_1=1$, so we invert Eq.~\eqref{eq:bin-mi} at the measured $I(N/2,N)$ to define an MI-equivalent cluster number $k_{\rm MI}$.
Throughout the inactive phase, $k_{\rm MI}-1$ decreases monotonically with system size and is consistent with $k_{\rm MI}\to 1$ as $N\to\infty$. Fig.~\ref{fig:mutual-info}(b) shows the finite-size data, plotted as $\Delta k_{\rm MI}=k_{\rm MI}-1$, at $p=0.55$ and $p=0.60$.
The cluster-count distribution $P(k)$, by contrast, shows that surviving configurations are frequently multi-cluster: $P(k)$ develops an approximately exponential tail in $k$ whose weight grows toward $p_c$, so that away from the deep inactive regime a typical configuration contains several clusters\,(see Appendix~\ref{sec:cluster-count}). Were these geometrically distinct clusters positioned independently, $k$ of them would contribute $I_k=H[\mathrm{Bin}(k,1/2)]>1$~bit and $k_{\rm MI}$ would rise above unity accordingly; instead $k_{\rm MI}\to 1$. That the bipartite information collapses to a single bit despite this recurrent multi-cluster geometry is the information-theoretic signature of single flock behavior.
The clusters do not move independently: they share a single left-right positional degree of freedom --- that of the enclosing flock --- and therefore behave as components of one bound object rather than as independent clusters.
The active phase is qualitatively different: there $\bar k$ grows with $N$ because many active domains are separated by holes, yet the mutual information remains small and nearly $N$-independent [inset of Fig.~\ref{fig:mutual-info}(a)], consistent with a bulk-like state lacking any extended positional degree of freedom.

\section{Discussion and conclusion}
\label{sec:discussion}
In this work, we characterized the quasi-stationary distribution of the bond directed-percolation line of the Domany--Kinzel automaton at the level of its full conditional distribution, using a matrix-product-state representation of the QSD obtained by power iteration of the projected transfer matrix to evaluate bipartite-information diagnostics on chains up to $N=1024$.
The central finding is that the inactive-phase QSD behaves as a single bound object even when its geometric appearance is fragmented into several clusters threaded by holes.
Two distribution-level signatures establish this conclusion independently: the bond-resolved mutual information takes the universal form of Eq.~\eqref{eq:mi-scaling}, whose leading term is exactly the one bit produced by a single uniformly delocalized position; and the finite-size scaling $k_{\rm MI}\to 1$ persists even though the surviving configurations are recurrently multi-cluster, the cluster count $k$ retaining an exponential tail well beyond $k=1$.
A picture of multiple independently wandering clusters is incompatible with either observation; together, the two identify the inactive-phase QSD with the flock scenario.

This refines, at the level of the full conditional distribution, the bound-survivor picture familiar from rigorous QSD analyses of related absorbing-state systems~\cite{Ferrari1995,Ferrari1996,Champagnat2016,ChetriteTouchette2015,Collet2013}: what survives the conditioning is not a single cluster but a flock of variable internal filling, with geometrically distinct clusters tied together into a single collective position.
The signature is carried by the bipartite Shannon mutual information of the QSD --- an observable that, to our knowledge, has not previously been used as a diagnostic for absorbing-state QSDs, and that is structurally inaccessible to the moment-based methods that dominate quasi-stationary simulation~\cite{Dickman2002,Oliveira2005}.

Three caveats should be noted.
The chains are open with $N\le 1024$, so we do not access the QSD asymptotically near $p_c$ where finite-size scaling becomes non-monotone for the boundary conditions used here.
The flock characterization is geometric, in the sense that $\ell_{\rm act}$ and $\rho_{\rm flock}$ measure spatial extent and filling, not the underlying causal-genealogical bonds that motivate the ``single-cluster'' reading.
And the consistency of $I(\ell, N)\simeq h(\ell/N)$ with a uniformly delocalized flock coordinate does not, on its own, exclude more elaborate spatial-position distributions with the same bipartite-MI profile; direct measurement of the cluster position distribution would close this remaining gap.

We believe that our framework that underlies this analysis~\cite{Derrida1993,Johnson2010,Banuls2019,Helms2019}  naturally extends to entanglement-spectrum diagnostics of the projected transfer matrix, to effective-Hamiltonian and large-deviation descriptions of absorbing-state QSDs, and to other one-dimensional absorbing-state models where conditioning on survival may produce analogous bound structures.

\textit{Note added.}---While finalizing this manuscript, we became aware of a related preprint~\cite{Boesl2026}, which interprets the entanglement transition in the $(2+0)$D quantum state obtained from the Domany--Kinzel automaton. Their results complement ours, providing a quantum-state perspective on the directed-percolation entanglement structure.

\section*{Acknowledgements}
H.-Y.L. acknowledges the hospitality of the Institute for Solid State Physics, The University of Tokyo, where part of this work was carried out.

\paragraph{Funding information}
K.H. was supported by JSPS KAKENHI Grant No.~24K06886. N.K.'s work is supported by JSPS KAKENHI Grant No.~23K25789. H.-Y.L was supported by the Basic Science Research Program through the National Research Foundation of Korea funded by the Ministry of Science and ICT [Grant No. RS-2023-00220471, RS-2025-16064392].

\begin{appendix}

\section{Local kernels and the two-layer transfer matrix}
\label{sec:S1}\label{app:kernels}

\subsection{DK update as a sum over parents}

The DK automaton evolves binary occupations $x_i(t)\in\{0,1\}$ on a chain of $N$ sites. A site at the next layer is active with a probability $P[n]$ that depends on the number $n\in\{0,1,2\}$ of active parents:
\begin{equation}
P[0]=0,\qquad P[1]=p_1,\qquad P[2]=p_2.
\label{eq:S-DK-rule}
\end{equation}
The bond-DP line is $p_1=p$, $p_2=p(2-p)$. The condition $P[0]=0$ makes $|0^N\rangle$ absorbing.

One time step of the DK update on a tilted square lattice consists of two sub-updates separated by a half-time-step. Sub-step $a$ generates an intermediate layer of $N{-}1$ sites at half-integer positions, each fed by two parents at integer positions on the input layer. Sub-step $b$ then generates the output layer of $N$ sites, with the bulk sites fed by two intermediate parents and the two boundary sites fed by a single intermediate parent (the chain is open). The two sub-steps return the chain to the original sublattice; one DK time step is therefore one ``two-layer'' transfer matrix.

\subsection{Local tensors \texorpdfstring{$\Wgate$, $\Vgate$, $\Ctensor$}{W, V, C}}

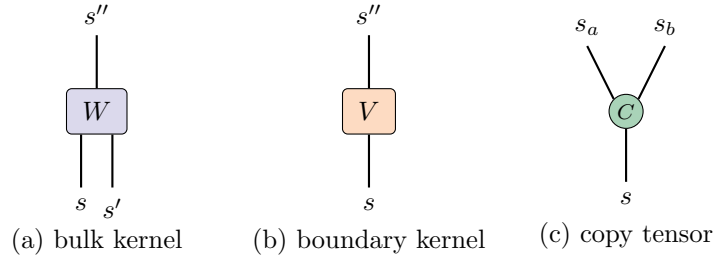
\begin{figure}[h]
\centering
\begin{tikzpicture}
\node[wtensor] (W) at (0,0) {$\Wgate$};
\draw[leg] ([xshift=2mm]W.south west) -- ++(0,-0.7) node[below,font=\small] {$s$};
\draw[leg] ([xshift=-2mm]W.south east) -- ++(0,-0.7) node[below,font=\small] {$s'$};
\draw[leg] (W.north) -- ++(0,0.7) node[above,font=\small] {$s''$};
\node[below=1.1cm of W, font=\small] {(a) bulk kernel};

\node[vtensor] (V) at (3.6,0) {$\Vgate$};
\draw[leg] (V.south) -- ++(0,-0.7) node[below,font=\small] {$s$};
\draw[leg] (V.north) -- ++(0,0.7) node[above,font=\small] {$s''$};
\node[below=1.1cm of V, font=\small] {(b) boundary kernel};

\node[ctensor] (C) at (7.0,0) {$\Ctensor$};
\draw[leg] (C.south) -- ++(0,-0.7) node[below,font=\small] {$s$};
\draw[leg] (C.north west) -- ++(-0.35,0.7) node[above,font=\small] {$s_a$};
\draw[leg] (C.north east) -- ++(0.35,0.7) node[above,font=\small] {$s_b$};
\node[below=1.1cm of C, font=\small] {(c) copy tensor};
\end{tikzpicture}
\caption{Local tensors of the DK update [Eqs.~\eqref{eq:S-W}--\eqref{eq:S-C}], in the convention of Ref.~\cite{Harada2019}. (a) Bulk three-leg kernel $\Wgate[s,s';s'']$ implements the two-parent rule $P[s{+}s']$. (b) Two-leg boundary kernel $\Vgate[s;s'']$ implements the single-parent rule $P[s]$. (c) Three-leg copy tensor $\Ctensor[s;s_a,s_b]=\delta_{s,s_a}\delta_{s,s_b}$ duplicates a site value into two outgoing wires; it sits at every internal site that feeds two adjacent gates.}
\label{fig:S-WV-tensors}
\end{figure}

Sub-steps $a$ and $b$ are built from three local tensors, following the convention of Ref.~\cite{Harada2019}. The bulk three-leg kernel $\Wgate$,
\begin{equation}
\Wgate[s,s';\,s''] \equiv (1-s'')\,(1-P[s{+}s']) + s''\,P[s{+}s'],
\label{eq:S-W}
\end{equation}
takes two parent occupations $s,s'\in\{0,1\}$ on a lower bond and returns the occupation $s''$ of the child site bridging them. With Eq.~\eqref{eq:S-DK-rule},
\begin{equation}
\Wgate[0,0;1]=0,\quad \Wgate[0,1;1]=\Wgate[1,0;1]=p_1,\quad \Wgate[1,1;1]=p_2,
\label{eq:S-Wentries}
\end{equation}
and $\Wgate[\cdot,\cdot;0]=1-\Wgate[\cdot,\cdot;1]$. The two-leg boundary kernel $\Vgate$ handles a child with only one parent (at the two ends of sub-step $b$):
\begin{equation}
\Vgate[s;\,s''] \equiv (1-s'')\,(1-P[s]) + s''\,P[s].
\label{eq:S-V}
\end{equation}
Whenever a site value is needed as an input to two adjacent gates, the tensor network represents the duplication explicitly through a three-leg copy (delta) tensor
\begin{equation}
\Ctensor[s;\,s_a,s_b] \;\equiv\; \delta_{s,s_a}\,\delta_{s,s_b},
\label{eq:S-C}
\end{equation}
which forces all three legs to carry the same binary value. Equations \eqref{eq:S-W}--\eqref{eq:S-C} reproduce the local tensors $W$, $V$, $C$ of the DK tensor network of Ref.~\cite{Harada2019}. We use the diagrammatic conventions of Fig.~\ref{fig:S-WV-tensors}.

\subsection{Sub-step \texorpdfstring{$a$}{a}: layer 1 (\texorpdfstring{$N\to N{-}1$}{N to N-1})}

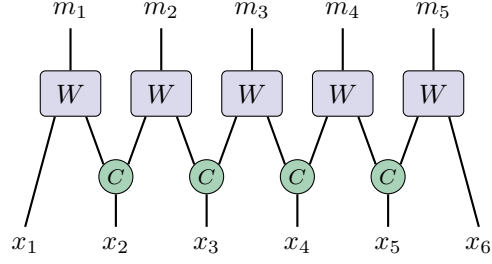
\begin{figure}[h]
\centering
\begin{tikzpicture}[x=1.2cm,y=1.0cm]
\foreach \i in {1,...,6} {\node[font=\small] (x\i) at (\i,-1.4) {$x_{\i}$};}
\foreach \i in {2,3,4,5} {\node[ctensor] (C\i) at (\i,-0.5) {$\Ctensor$};}
\foreach \i in {1,...,5} {%
  \pgfmathsetmacro{\xpos}{\i+0.5}
  \node[wtensor] (W\i) at (\xpos,0.6) {$\Wgate$};
}
\foreach \i in {1,...,5} {%
  \pgfmathsetmacro{\xpos}{\i+0.5}
  \node[font=\small] (m\i) at (\xpos,1.7) {$m_{\i}$};
}
\foreach \i in {2,3,4,5} {\draw[leg] (x\i) -- (C\i.south);}
\draw[leg] (x1.north) -- ([xshift=2mm]W1.south west);
\draw[leg] (x6.north) -- ([xshift=-2mm]W5.south east);
\draw[leg] (C2.north west) -- ([xshift=-2mm]W1.south east);
\draw[leg] (C2.north east) -- ([xshift=2mm]W2.south west);
\draw[leg] (C3.north west) -- ([xshift=-2mm]W2.south east);
\draw[leg] (C3.north east) -- ([xshift=2mm]W3.south west);
\draw[leg] (C4.north west) -- ([xshift=-2mm]W3.south east);
\draw[leg] (C4.north east) -- ([xshift=2mm]W4.south west);
\draw[leg] (C5.north west) -- ([xshift=-2mm]W4.south east);
\draw[leg] (C5.north east) -- ([xshift=2mm]W5.south west);
\foreach \i in {1,...,5} {\draw[leg] (W\i.north) -- (m\i);}
\end{tikzpicture}
\caption{Sub-step $a$ for $N=6$: brick wall of bulk gates $\Wgate$ producing the intermediate layer $(m_1,\ldots,m_5)$. Each internal input $x_k$ is duplicated by a $\Ctensor$ tensor and fed into two adjacent $\Wgate$ gates; the boundary inputs $x_1$ and $x_6$ feed a single gate each.}
\label{fig:S-layer1}
\end{figure}

Sub-step $a$ contracts $N{-}1$ copies of $\Wgate$ between adjacent pairs of input sites $(x_k,x_{k+1})$, producing the intermediate layer $(m_1,\ldots,m_{N-1})$:
\begin{equation}
T^{(a)}[x_1,\ldots,x_N;\,m_1,\ldots,m_{N-1}] = \prod_{k=1}^{N-1}\Wgate[x_k,x_{k+1};\,m_k].
\label{eq:S-Ta}
\end{equation}
Each internal input site $x_k$ ($k=2,\ldots,N{-}1$) is shared by gates $\Wgate_{k-1}$ and $\Wgate_k$ and is therefore duplicated through a $\Ctensor$ tensor; the boundary sites $x_1$ and $x_N$ feed only $\Wgate_1$ and $\Wgate_{N-1}$ respectively, and require no copy. The resulting tensor network is a one-row brick wall (Fig.~\ref{fig:S-layer1}).

\subsection{Sub-step \texorpdfstring{$b$}{b}: layer 2 (\texorpdfstring{$N{-}1\to N$}{N-1 to N})}

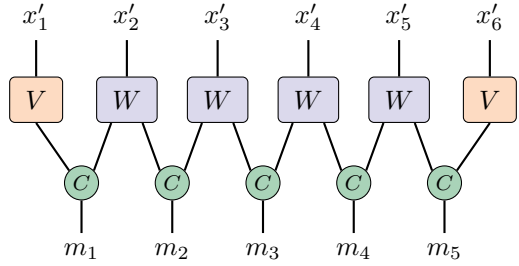
\begin{figure}[h]
\centering
\begin{tikzpicture}[x=1.2cm,y=1.0cm]
\foreach \i in {1,...,5} {%
  \pgfmathsetmacro{\xpos}{\i+0.5}
  \node[font=\small] (m\i) at (\xpos,-1.4) {$m_{\i}$};
}
\foreach \i in {1,...,5} {%
  \pgfmathsetmacro{\xpos}{\i+0.5}
  \node[ctensor] (C\i) at (\xpos,-0.5) {$\Ctensor$};
}
\node[vtensor] (Vl) at (1,0.6) {$\Vgate$};
\node[vtensor] (Vr) at (6,0.6) {$\Vgate$};
\foreach \i in {1,...,4} {%
  \pgfmathsetmacro{\xpos}{\i+1}
  \node[wtensor] (W\i) at (\xpos,0.6) {$\Wgate$};
}
\foreach \i in {1,...,6} {\node[font=\small] (xp\i) at (\i,1.7) {$x'_{\i}$};}
\foreach \i in {1,...,5} {\draw[leg] (m\i) -- (C\i.south);}
\draw[leg] (C1.north west) -- (Vl.south);
\draw[leg] (C1.north east) -- ([xshift=2mm]W1.south west);
\draw[leg] (C2.north west) -- ([xshift=-2mm]W1.south east);
\draw[leg] (C2.north east) -- ([xshift=2mm]W2.south west);
\draw[leg] (C3.north west) -- ([xshift=-2mm]W2.south east);
\draw[leg] (C3.north east) -- ([xshift=2mm]W3.south west);
\draw[leg] (C4.north west) -- ([xshift=-2mm]W3.south east);
\draw[leg] (C4.north east) -- ([xshift=2mm]W4.south west);
\draw[leg] (C5.north west) -- ([xshift=-2mm]W4.south east);
\draw[leg] (C5.north east) -- (Vr.south);
\draw[leg] (Vl.north) -- (xp1);
\draw[leg] (Vr.north) -- (xp6);
\foreach \i in {1,...,4} {%
  \pgfmathtruncatemacro{\out}{\i+1}
  \draw[leg] (W\i.north) -- (xp\out);
}
\end{tikzpicture}
\caption{Sub-step $b$ for $N=6$: each intermediate $m_k$ is duplicated by a $\Ctensor$ tensor and fed into two operators of layer 2. Boundary $\Vgate$ gates generate $x'_1$ from $m_1$ and $x'_N$ from $m_{N-1}$; bulk $\Wgate$ gates generate $x'_{k+1}$ from $(m_k,m_{k+1})$ for $k=1,\ldots,N{-}2$.}
\label{fig:S-layer2}
\end{figure}

Sub-step $b$ takes the intermediate layer $(m_1,\ldots,m_{N-1})$ to the output layer $(x'_1,\ldots,x'_N)$. Bulk output sites $x'_{k+1}$ for $k=1,\ldots,N{-}2$ are generated by a $\Wgate$ gate with parents $(m_k,m_{k+1})$, while the two boundary sites $x'_1$ and $x'_N$ have only one parent each and are generated by $\Vgate$ acting on $m_1$ and $m_{N-1}$ respectively:
\begin{equation}
T^{(b)}[m_1,\ldots,m_{N-1};\,x'_1,\ldots,x'_N]
= \Vgate[m_1;x'_1]\,\Vgate[m_{N-1};x'_N]\,\prod_{k=1}^{N-2}\Wgate[m_k,m_{k+1};\,x'_{k+1}].
\label{eq:S-Tb}
\end{equation}
Every intermediate $m_k$ is shared by two gates---$(\Vgate,\Wgate)$ at the left end, $(\Wgate,\Wgate)$ in the bulk, $(\Wgate,\Vgate)$ at the right end---and is therefore duplicated by a $\Ctensor$ tensor. The brick-wall pattern is shifted by half a unit relative to sub-step $a$ (Fig.~\ref{fig:S-layer2}).

\subsection{Two-step transfer matrix}

The full one-step DK stochastic transfer matrix is the contraction of sub-steps $a$ and $b$ over the intermediate layer:
\begin{equation}
T[x_1,\ldots,x_N;\,x'_1,\ldots,x'_N]
= \sum_{m_1,\ldots,m_{N-1}\in\{0,1\}}
T^{(b)}[m;x']\,T^{(a)}[x;m].
\label{eq:S-Tfull}
\end{equation}
Stacking the two brick walls of Figs.~\ref{fig:S-layer1} and \ref{fig:S-layer2} gives the $\Ctensor$-$\Wgate$-$\Vgate$ tensor network of Fig.~\ref{fig:S-2step}, built from the same three local tensors used in Fig.~2 of Ref.~\cite{Harada2019}. We verified Eq.~\eqref{eq:S-Tfull} against a brute-force enumeration of $T$ for $N\le 10$ at multiple values of $p$.

\paragraph{Open-boundary convention.}
Our construction places the intermediate row at half-integer positions and consequently has $N{-}1$ intermediate sites between two rows of $N$ physical sites; the two boundary $\Vgate$ kernels both belong to sub-step $b$. Reference~\cite{Harada2019} draws the same DK tensor network on the tilted lattice with every row carrying $N$ sites, in which case a single $\Vgate$ kernel appears at one end of sub-step $a$ and another $\Vgate$ kernel at the opposite end of sub-step $b$. The two conventions correspond to slightly different open-boundary versions of the bond-DP DK chain that agree in the bulk and differ only by an $O(1)$ boundary correction; on the chains $N\ge 128$ studied here the difference is well within the finite-size scatter of every observable reported in the main text. We adopt the present convention because it matches the implementation in which the intermediate row is generated directly by sub-step $a$ with no boundary kernels.

\section{MPS representation and gate-by-gate application}
\label{sec:S2}\label{app:mps}

\subsection{Probability MPS and the flat covector}

A probability distribution $P(x_1,\ldots,x_N)$ over $\{0,1\}^N$ is represented as an MPS,
\begin{equation}
|P\rangle = \sum_{x_1,\ldots,x_N} P(x_1,\ldots,x_N)\,|x_1,\ldots,x_N\rangle
= \sum_{\{x_i\}}\mathrm{Tr}\big(A^{[1]}_{x_1}A^{[2]}_{x_2}\cdots A^{[N]}_{x_N}\big)\,|x_1\cdots x_N\rangle,
\label{eq:S-MPS}
\end{equation}
with site tensors $A^{[k]}_{x_k}\in\mathbb R^{\chi_{k-1}\times\chi_k}$ and bond dimensions $\chi_k$. The crucial distinction from the wavefunction MPS of Ref.~\cite{Harada2019}---where the amplitudes are $\exp[H_2/2]\,P(X)$ and the state is normalized by the Born $L^2$ inner product so that the second-order R\'enyi entropy enters the wavefunction normalization---is that here $|P\rangle$ stores the probability itself, and physical averages are evaluated with the \emph{flat} covector
\begin{equation}
\langle\onevec| \;=\; \sum_{x_1,\ldots,x_N} \langle x_1\cdots x_N|,\qquad
\langle\onevec|P\rangle \;=\; \sum_X P(X) \;=\; 1.
\label{eq:S-flat}
\end{equation}
We therefore replace the $L^2$ Born-rule contraction of a wavefunction MPS by a contraction with the flat covector wherever a probability sum is required. The tensor network for $T$ (Fig.~\ref{fig:S-2step}) is identical in the two conventions, since $T$ is linear; only the normalization of the state and the choice of left contractor differ.

\subsection{Gate application}

\begin{figure}[h]
\centering
\begin{tikzpicture}[x=1.2cm,y=1.0cm]
\node[mpstensor] (Ak) at (0,0) {$A^{[k]}$};
\node[mpstensor] (Akp) at (1.7,0) {$A^{[k\!+\!1]}$};
\draw[bond] (Ak.east) -- (Akp.west);
\draw[bond] (Ak.west) -- ++(-0.6,0);
\draw[bond] (Akp.east) -- ++(0.6,0);
\draw[leg] (Ak.south) -- ++(0,-0.5);
\draw[leg] (Akp.south) -- ++(0,-0.5);
\node[wtensor] (W) at (0.85,1.1) {$\Wgate$};
\draw[leg] (Ak.north) -- ([xshift=2mm]W.south west);
\draw[leg] (Akp.north) -- ([xshift=-2mm]W.south east);
\draw[leg] (W.north) -- ++(-0.25,0.5);
\draw[leg] (W.north) -- ++(0.25,0.5);
\node at (3.5,0.5) {$\xrightarrow{\;\text{SVD}\;}$};
\node[mpstensor] (At) at (5.0,0.5) {$\tilde A^{[k]}$};
\node[mpstensor] (Atp) at (6.7,0.5) {$\tilde A^{[k\!+\!1]}$};
\draw[bond] (At.east) -- (Atp.west) node[midway,above,font=\scriptsize] {$\le\chi_{\max}$};
\draw[bond] (At.west) -- ++(-0.6,0);
\draw[bond] (Atp.east) -- ++(0.6,0);
\draw[leg] (At.north) -- ++(0,0.6);
\draw[leg] (Atp.north) -- ++(0,0.6);
\end{tikzpicture}
\caption{Gate-by-gate application: a two-site block $A^{[k]}A^{[k+1]}\cdot\Wgate$ is reshaped and SVD-decomposed back into MPS form with the truncation $\chi\le\chi_{\max}$.}
\label{fig:S-gate-apply}
\end{figure}
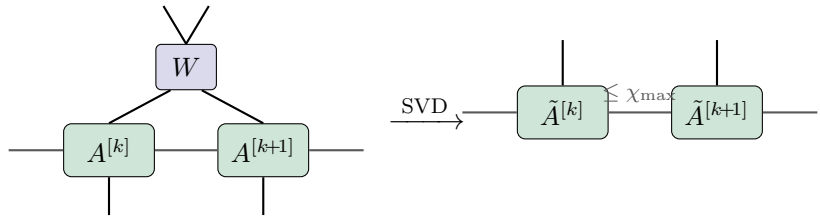

A bulk gate $\Wgate$ couples two adjacent sites; applied to a contiguous pair $(A^{[k]},A^{[k+1]})$ it produces a two-site block which is then refactorized by a singular-value decomposition (SVD) into updated site tensors $\tilde A^{[k]}$, $\tilde A^{[k+1]}$ (Fig.~\ref{fig:S-gate-apply}). Truncation keeps the leading $\chi_{\max}$ singular values, discarding those below an absolute cutoff $\varepsilon_{\rm cut}$. Boundary $\Vgate$ gates act on a single site and require no refactorization.

\subsection{Sweep order and bond-dimension growth}

Sub-step $a$ is applied as a left-to-right sweep of $\Wgate$ gates; the gates do not commute pairwise, but the brick-wall layout of Fig.~\ref{fig:S-layer1} fixes a unique application order and the sweep is exact up to the truncation of the SVD. Sub-step $b$ is applied right-to-left, starting from the right-boundary $\Vgate$, sweeping through the bulk $\Wgate$ gates, and closing with the left-boundary $\Vgate$. Each gate application is followed by an immediate SVD recompression; we use $\chi_{\max}=120$ and $\varepsilon_{\rm cut}=10^{-15}$ throughout. The bond dimension is monitored along the sweep; deep in the inactive phase the post-truncation $\chi$ saturates well below $\chi_{\max}$ on chains up to $N=1024$.

In the implementation, the MPS is kept at fixed length $N$ throughout both sub-steps by absorbing each $\Wgate$ gate together with a pass-through identity on the unused physical leg, so that sub-step $a$'s two-site gate at bond $(k,k{+}1)$ reads $T^{(a)}_{kk{+}1}[s'_k,s'_{k+1};s_k,s_{k+1}]=\Wgate[s_k,s_{k+1};s'_k]\,\delta_{s_{k+1},s'_{k+1}}$ (left site updated to $m_k$, right site passed through), and sub-step $b$'s bulk gate at $(k,k{+}1)$ reads $T^{(b)}_{kk{+}1}[s'_k,s'_{k+1};m_k,m_{k+1}]=\delta_{m_k,s'_k}\,\Wgate[m_k,m_{k+1};s'_{k+1}]$ (left site passed through, right site updated to $x'_{k+1}$). The two boundary $\Vgate$ contributions of sub-step $b$ are absorbed into the two terminal two-site gates. This pass-through trick avoids changing the chain length to $N{-}1$ between sub-steps; it leaves $T$ unchanged.

\section{Projection and power iteration}
\label{sec:S3}\label{app:projection}

\subsection{Projector and projected transfer matrix}

The QSD $|P_{\rm QS}\rangle$ is the leading non-negative right eigenvector of the projected transfer matrix
\begin{equation}
\Pi T\Pi,\qquad \Pi = \mathbbm 1 - |0^N\rangle\langle 0^N|,
\label{eq:S-projector}
\end{equation}
with eigenvalue $\lambda_1<1$ equal to the survival probability per DK step. We implement $\Pi|P\rangle$ as a rank-one subtraction at the MPS level:
\begin{equation}
\Pi|P\rangle = |P\rangle - P(0^N)\,|0^N\rangle,\qquad P(0^N)=\langle 0^N|P\rangle.
\label{eq:S-projection}
\end{equation}
The product state $|0^N\rangle$ is a $\chi=1$ MPS, and $\langle 0^N|P\rangle$ is the single-amplitude contraction of $|P\rangle$ with the projector onto $x_k=0$ at every site---an $O(N\chi^2)$ contraction. The subtraction itself nominally raises the bond dimension by one, which is reabsorbed by the next SVD recompression.

\subsection{Power iteration}

The QSD is reached by power iteration on Eq.~\eqref{eq:S-projector}: starting from a non-absorbing initial MPS, we repeatedly apply the two-layer transfer matrix $T$ (Sec.~\ref{sec:S2}), project out the absorbing component via Eq.~\eqref{eq:S-projection}, and renormalize so that $\langle\onevec|P\rangle=1$. The dominant eigenvalue $\lambda_1$ is read off from the renormalization constants between successive iterations:
\begin{equation}
\lambda_1 \;=\; \lim_{t\to\infty}\frac{\langle\onevec|\,\Pi T\,|P_t\rangle}{\langle\onevec|P_t\rangle}.
\label{eq:S-lambda1}
\end{equation}
The full per-step update is summarized as Algorithm~\ref{alg:S-power}.

\begin{algorithm}[H]
\caption{One step of the QSD power iteration.}
\label{alg:S-power}
\KwIn{probability MPS $|P_t\rangle$ with $\langle\onevec|P_t\rangle=1$; kernel data $(p_1,p_2)$; bond cap $\chi_{\max}$; cutoff $\varepsilon_{\rm cut}$.}
Apply sub-step $a$: sweep $\Wgate$ gates left-to-right (no truncation)\;
Apply sub-step $b$: right-boundary $\Vgate$, bulk $\Wgate$ gates right-to-left, left-boundary $\Vgate$; SVD-truncate at $(\chi_{\max},\varepsilon_{\rm cut})$ after each gate\;
Compute $w_0 \gets \langle 0^N|P\rangle$\;
$|P\rangle \gets |P\rangle - w_0\,|0^N\rangle$\tcp*{$\Pi$-projection}
Compute $Z \gets \langle\onevec|P\rangle$; set $\lambda_1^{(t)} \gets Z$\;
$|P\rangle \gets |P\rangle/Z$\;
\KwOut{$|P_{t+1}\rangle$ and $\lambda_1^{(t)}$.}
\end{algorithm}

\subsection{Convergence diagnostics}

Two independent criteria gate convergence. (i) The overlap between successive iterates, $1-|\langle P_t|P_{t-\Delta t}\rangle|$ (after two-norm normalization), is required to fall below $10^{-6}$; this controls the structural convergence of the MPS. (ii) The eigenvalue estimate $\lambda_1^{(t)}$ is required to vary by less than $10^{-4}$ between successive iterates; this provides an independent stability check that is sensitive deep in the inactive phase, where $\lambda_1$ approaches 1 only slowly.

\section{Perfect sampling}
\label{sec:S4}\label{app:sampling}

\subsection{Sequential conditional sampling}

Exact, rejection-free samples from $P_{\rm QS}$ are obtained by sequentially drawing each site from its left-conditional distribution and updating the conditioning. For a probability MPS the conditional probability of $x_k$ given a fixed prefix $x_1,\ldots,x_{k-1}$ is
\begin{equation}
P(x_k\mid x_1,\ldots,x_{k-1})
= \frac{\sum_{x_{k+1},\ldots,x_N} P(x_1,\ldots,x_N)}{\sum_{x_{k},\ldots,x_N} P(x_1,\ldots,x_N)}
= \frac{L^{[k-1]}\cdot A^{[k]}_{x_k}\cdot R^{[k+1]}}{\sum_{x'_k\in\{0,1\}} L^{[k-1]}\cdot A^{[k]}_{x'_k}\cdot R^{[k+1]}},
\label{eq:S-cond}
\end{equation}
where $L^{[k-1]}$ is the left environment built from the already-fixed prefix and $R^{[k+1]}$ is the right environment built by flat-summation over the un-fixed suffix.

\subsection{Right environment caching}

\begin{figure}[h]
\centering
\begin{tikzpicture}[x=1.0cm,y=0.9cm]
\node[mpstensor] (A1) at (1.5,0) {$A^{[1]}$};
\node[mpstensor] (A2) at (3.0,0) {$A^{[2]}$};
\node[mpstensor] (A3) at (4.5,0) {$A^{[3]}$};
\node[mpstensor] (A4) at (6.0,0) {$A^{[4]}$};
\node at (7.3,0) {$\cdots$};
\node[mpstensor] (AN) at (8.5,0) {$A^{[N]}$};
\draw[bond] (A1.east) -- (A2.west);
\draw[bond] (A2.east) -- (A3.west);
\draw[bond] (A3.east) -- (A4.west);
\draw[bond] (A4.east) -- ++(0.5,0);
\draw[bond] (AN.west) -- ++(-0.5,0);
\draw[bond] (A1.west) -- ++(-0.5,0);
\draw[bond] (AN.east) -- ++(0.5,0);
\node[flatvec] (f1) at (1.5,-1.1) {$\onevec$};
\node[flatvec] (f2) at (3.0,-1.1) {$\onevec$};
\node[flatvec] (f3) at (4.5,-1.1) {$\onevec$};
\node[flatvec] (f4) at (6.0,-1.1) {$\onevec$};
\node[flatvec] (fN) at (8.5,-1.1) {$\onevec$};
\draw[leg] (A1.south) -- (f1.north);
\draw[leg] (A2.south) -- (f2.north);
\draw[leg] (A3.south) -- (f3.north);
\draw[leg] (A4.south) -- (f4.north);
\draw[leg] (AN.south) -- (fN.north);
\node[font=\small] (R1) at (0.7,-2.1) {$R^{[1]}$};
\node[font=\small] (R2) at (2.25,-2.1) {$R^{[2]}$};
\node[font=\small] (R5) at (6.7,-2.1) {$R^{[5]}$};
\draw[->,>=Stealth] (R5) -- (R2);
\draw[->,>=Stealth] (R2) -- (R1);
\end{tikzpicture}
\caption{Right environments. Each site tensor is contracted with a flat vector $\onevec=(1,1)$ on the physical leg, and the resulting two-leg matrices are accumulated right-to-left, $R^{[k]} = (A^{[k]}\cdot\onevec_k)\,R^{[k+1]}$, starting from the trivial $R^{[N+1]}=1$.}
\label{fig:S-rightenv}
\end{figure}
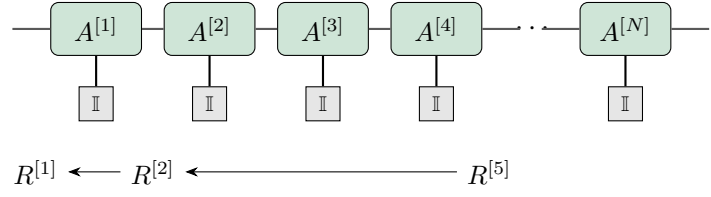

The right environments
\begin{equation}
R^{[k]} \;=\; \sum_{x_k,\ldots,x_N} A^{[k]}_{x_k}A^{[k+1]}_{x_{k+1}}\cdots A^{[N]}_{x_N}
\label{eq:S-rightenv}
\end{equation}
do not depend on the sample drawn so far and can be pre-computed once per MPS by a right-to-left sweep, contracting each site tensor with the flat covector $\langle\onevec_k|=(1,1)$ on the physical leg (Fig.~\ref{fig:S-rightenv}). Storing the full set $\{R^{[k]}\}_{k=1}^{N+1}$ requires $O(N\chi^2)$ memory, after which each subsequent sample costs $O(N\chi^2)$.

\subsection{Site-by-site draw}

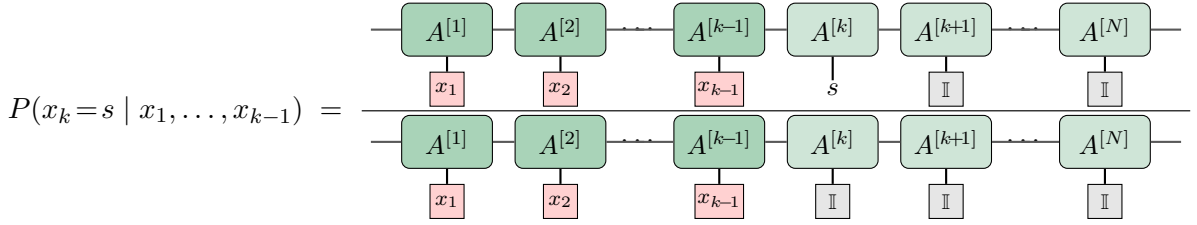
\begin{figure}[h]
\centering
\[
P(x_k\!=\!s\mid x_1,\ldots,x_{k-1}) \;=\;
\frac{\;
\vcenter{\hbox{\begin{tikzpicture}[x=1.0cm,y=0.75cm]
\node[mpstensor,fill=ForestGreen!35] (A1) at (1.0,0) {$A^{[1]}$};
\node[mpstensor,fill=ForestGreen!35] (A2) at (2.5,0) {$A^{[2]}$};
\node at (3.55,0) {$\cdots$};
\node[mpstensor,fill=ForestGreen!35] (Akm) at (4.6,0) {$A^{[k\!-\!1]}$};
\node[mpstensor] (Ak) at (6.1,0) {$A^{[k]}$};
\node[mpstensor] (Akp) at (7.6,0) {$A^{[k\!+\!1]}$};
\node at (8.65,0) {$\cdots$};
\node[mpstensor] (AN) at (9.7,0) {$A^{[N]}$};
\draw[bond] (A1.east) -- (A2.west);
\draw[bond] (A2.east) -- ++(0.5,0);
\draw[bond] (Akm.west) -- ++(-0.5,0);
\draw[bond] (Akm.east) -- (Ak.west);
\draw[bond] (Ak.east) -- (Akp.west);
\draw[bond] (Akp.east) -- ++(0.5,0);
\draw[bond] (AN.west) -- ++(-0.5,0);
\draw[bond] (A1.west) -- ++(-0.4,0);
\draw[bond] (AN.east) -- ++(0.4,0);
\node[proj] (p1) at (1.0,-1.05) {$x_1$};
\node[proj] (p2) at (2.5,-1.05) {$x_2$};
\node[proj] (pkm) at (4.6,-1.05) {$x_{k\!-\!1}$};
\node[font=\small,inner sep=1pt] (sk) at (6.1,-1.05) {$s$};
\node[flatvec] (fkp) at (7.6,-1.05) {$\onevec$};
\node[flatvec] (fN) at (9.7,-1.05) {$\onevec$};
\draw[leg] (A1.south) -- (p1.north);
\draw[leg] (A2.south) -- (p2.north);
\draw[leg] (Akm.south) -- (pkm.north);
\draw[leg] (Ak.south) -- (sk.north);
\draw[leg] (Akp.south) -- (fkp.north);
\draw[leg] (AN.south) -- (fN.north);
\end{tikzpicture}}}
\;}{\;
\vcenter{\hbox{\begin{tikzpicture}[x=1.0cm,y=0.75cm]
\node[mpstensor,fill=ForestGreen!35] (A1) at (1.0,0) {$A^{[1]}$};
\node[mpstensor,fill=ForestGreen!35] (A2) at (2.5,0) {$A^{[2]}$};
\node at (3.55,0) {$\cdots$};
\node[mpstensor,fill=ForestGreen!35] (Akm) at (4.6,0) {$A^{[k\!-\!1]}$};
\node[mpstensor] (Ak) at (6.1,0) {$A^{[k]}$};
\node[mpstensor] (Akp) at (7.6,0) {$A^{[k\!+\!1]}$};
\node at (8.65,0) {$\cdots$};
\node[mpstensor] (AN) at (9.7,0) {$A^{[N]}$};
\draw[bond] (A1.east) -- (A2.west);
\draw[bond] (A2.east) -- ++(0.5,0);
\draw[bond] (Akm.west) -- ++(-0.5,0);
\draw[bond] (Akm.east) -- (Ak.west);
\draw[bond] (Ak.east) -- (Akp.west);
\draw[bond] (Akp.east) -- ++(0.5,0);
\draw[bond] (AN.west) -- ++(-0.5,0);
\draw[bond] (A1.west) -- ++(-0.4,0);
\draw[bond] (AN.east) -- ++(0.4,0);
\node[proj] (p1) at (1.0,-1.05) {$x_1$};
\node[proj] (p2) at (2.5,-1.05) {$x_2$};
\node[proj] (pkm) at (4.6,-1.05) {$x_{k\!-\!1}$};
\node[flatvec] (fk) at (6.1,-1.05) {$\onevec$};
\node[flatvec] (fkp) at (7.6,-1.05) {$\onevec$};
\node[flatvec] (fN) at (9.7,-1.05) {$\onevec$};
\draw[leg] (A1.south) -- (p1.north);
\draw[leg] (A2.south) -- (p2.north);
\draw[leg] (Akm.south) -- (pkm.north);
\draw[leg] (Ak.south) -- (fk.north);
\draw[leg] (Akp.south) -- (fkp.north);
\draw[leg] (AN.south) -- (fN.north);
\end{tikzpicture}}}
\;}
\]
\caption{Diagrammatic form of Eq.~\eqref{eq:S-cond}. The conditional probability is the ratio of two MPS contractions that differ only at site $k$: the numerator leaves site $k$ open at the trial value $s$ (this is the weight $w_k(s)$ of Eq.~\eqref{eq:S-weights}), while the denominator contracts site $k$ with the flat covector $\onevec=(1,1)$ and equals $\sum_{s'}w_k(s')$. Prefix sites $1,\ldots,k{-}1$ are projected onto the already-drawn values $x_1,\ldots,x_{k-1}$ (giving the left environment $L^{[k-1]}$); suffix sites $k{+}1,\ldots,N$ are contracted with the flat covector (giving the cached right environment $R^{[k+1]}$).}
\label{fig:S-cond-diag}
\end{figure}

Given the precomputed $\{R^{[k]}\}$, the left environment is initialized as $L^{[0]}=1$. At each site $k=1,\ldots,N$ the two unnormalized scalars
\begin{equation}
w_k(s) \;=\; L^{[k-1]}\cdot A^{[k]}_{s}\cdot R^{[k+1]},\qquad s\in\{0,1\},
\label{eq:S-weights}
\end{equation}
are evaluated (Fig.~\ref{fig:S-cond-diag}); $x_k$ is drawn from the categorical distribution $w_k(s)/(w_k(0)+w_k(1))$ and the left environment is updated as $L^{[k]} \gets L^{[k-1]}\cdot A^{[k]}_{x_k}$. The algorithm is summarized as Algorithm~\ref{alg:S-sample}; the resulting sample $X=(x_1,\ldots,x_N)$ is drawn exactly from $P$ provided $|P\rangle$ is non-negative on physical configurations.

\begin{algorithm}[H]
\caption{Perfect sampling from a probability MPS.}
\label{alg:S-sample}
\KwIn{MPS $|P\rangle$ with site tensors $\{A^{[k]}\}_{k=1}^N$; precomputed right environments $\{R^{[k]}\}_{k=2}^{N+1}$.}
$L \gets 1$\;
\For{$k = 1,\ldots,N$}{
  \For{$s \in \{0,1\}$}{
    $w_k(s) \gets L\cdot A^{[k]}_s\cdot R^{[k+1]}$\;
  }
  Draw $x_k$ from the categorical distribution $w_k(s)/\sum_{s'}w_k(s')$\;
  $L \gets L\cdot A^{[k]}_{x_k}$; rescale $L\gets L/\|L\|$ for numerical stability\;
}
\KwOut{sample $X=(x_1,\ldots,x_N)$.}
\end{algorithm}

\subsection{Correctness}

Telescoping the per-step ratios in Eq.~\eqref{eq:S-cond} gives
\begin{equation}
\prod_{k=1}^N P(x_k\mid x_1,\ldots,x_{k-1})
= \frac{P(x_1,\ldots,x_N)}{\sum_X P(X)}
= P(x_1,\ldots,x_N),
\end{equation}
so the joint draw from Algorithm~\ref{alg:S-sample} is exactly distributed according to $P$. No thermalization or rejection step is used; the only source of error is the SVD-truncation error of $|P\rangle$ itself.

\section{Convergence of the projected eigenvalue}
\label{sec:S6}\label{app:eigenvalue}

\begin{figure}[h]
\centering
\includegraphics[width=0.4\linewidth]{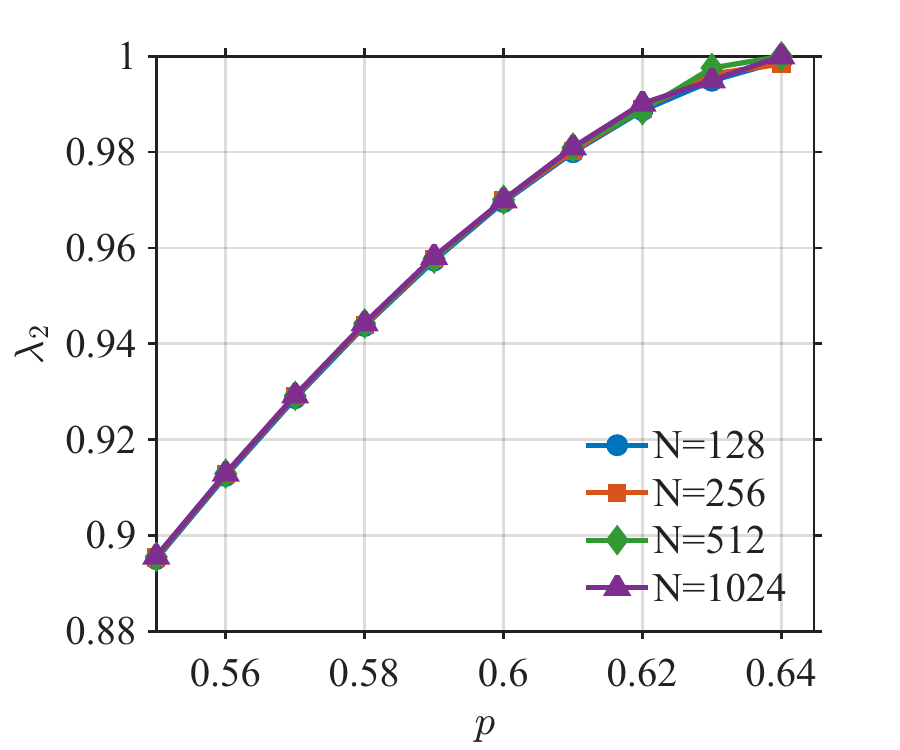}
\caption{Converged projected dominant eigenvalue $\lambda_1$ versus $p$ in the inactive phase for $N=128,~256,~512,~1024$. The four curves collapse onto a common, $N$-independent limit across the inactive phase, consistent with the QSD being a thermodynamic-limit object on the chains studied.}
\label{fig:S-eigen}
\end{figure}

Figure~\ref{fig:S-eigen} shows the converged $\lambda_1(p)$ in the inactive phase for $N=128,256,512,1024$. Across the inactive phase $\lambda_1$ approaches unity from below as $p$ approaches $p_c$, and the four chain lengths overlap closely, confirming that the projected dominant eigenvalue---and hence the QSD itself---has settled to its thermodynamic-limit value on the chains used in the main text. We obtained these values by running Algorithm~\ref{alg:S-power} until both convergence criteria of Sec.~\ref{sec:S3} are satisfied.

The full bond-dimension and truncation parameters used throughout the main text are $\chi_{\max}=120$ and $\varepsilon_{\rm cut}=10^{-15}$; the post-truncation $\chi$ saturates well below $\chi_{\max}$ on all chains studied.

\section{Subcritical correlation length $\xi_\perp(p)$ and DP universality}
\label{sec:xi-perp}

In the inactive phase, the QSD admits a connected one-site density--density correlator
\begin{equation}
G(r) \;\equiv\; \langle x_i\,x_{i+r}\rangle_{\mathrm{QS}} \;-\; \langle x_i\rangle_{\mathrm{QS}}\langle x_{i+r}\rangle_{\mathrm{QS}},
\label{eq:Gr-def}
\end{equation}
evaluated in the bulk of the chain. For $r$ in the range $1\ll r\ll N$ the connected correlator decays exponentially,
\begin{equation}
G(r) \;\sim\; e^{-r/\xi_\perp(p)},
\label{eq:Gr-exp}
\end{equation}
which defines the subcritical correlation length $\xi_\perp(p)$ used throughout this work.

\begin{figure}[h]
\centering
\includegraphics[width=0.35\linewidth]{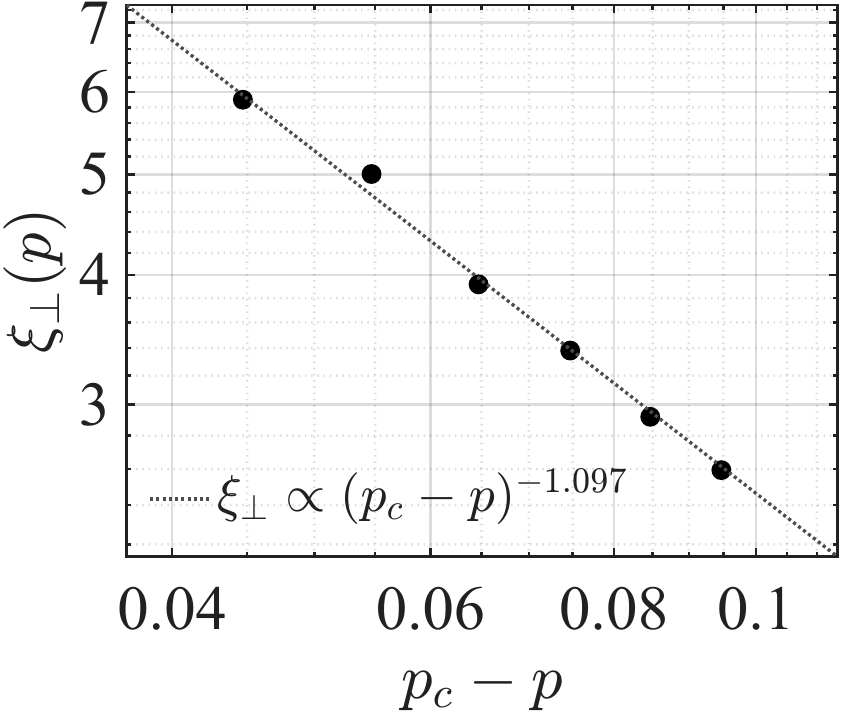}
\caption{Subcritical DP correlation length $\xi_\perp(p)$ extracted from the connected one-site density--density correlator of the QSD [Eq.~\eqref{eq:Gr-def}], on a log--log plot versus $p_c-p$. The dashed line is the power-law fit $\xi_\perp\propto(p_c-p)^{-\nu_\perp}$ with $\nu_\perp\simeq 1.097$, in agreement with the $(1{+}1)$D DP literature value~\cite{Hinrichsen2000,Jensen2004}.}
\label{fig:S-xi-perp}
\end{figure}

We extract $\xi_\perp(p)$ from a least-squares fit of $\log G(r)$ to a linear function on the window where the connected correlator dominates statistical noise and is not yet contaminated by the finite-size mean offset (typically $2\le r\le 0.4\,N$ at the chains we use). Figure~\ref{fig:S-xi-perp} shows the resulting $\xi_\perp(p)$ on a log--log plot against $p_c-p$. A power-law fit gives
\begin{equation}
\xi_\perp(p) \;\propto\; (p_c-p)^{-\nu_\perp}, \qquad \nu_\perp \;\simeq\; 1.097,
\label{eq:nu-perp}
\end{equation}
in agreement with the standard $(1{+}1)$-dimensional directed-percolation exponent $\nu_\perp\simeq 1.0968$~\cite{Hinrichsen2000,Jensen2004}.

Two consequences are relevant for the rest of the analysis. First, the $\xi_\perp(p)$ that enters the projected-iteration scheme as a measurable QSD observable scales with the universality-class exponent expected for $(1{+}1)$D directed percolation, providing a non-trivial consistency check that the projected dynamics captures the bona fide subcritical DP correlation structure rather than some boundary-induced or finite-size artefact. Second, the extracted values populate the range $\xi_\perp\in[2.5,\,6]$ over the $p$ window used in the main text, so the small parameter $\xi_\perp/N\le 6/128\simeq 5\%$ on the smallest chains and decreases as $1/N$ at fixed $p$, justifying the leading-order expansions of Sec.~\ref{sec:F-shape} below.

\section{Cluster-count statistics}
\label{sec:cluster-count}

\begin{figure}[h]
\centering
\includegraphics[width=0.99\linewidth]{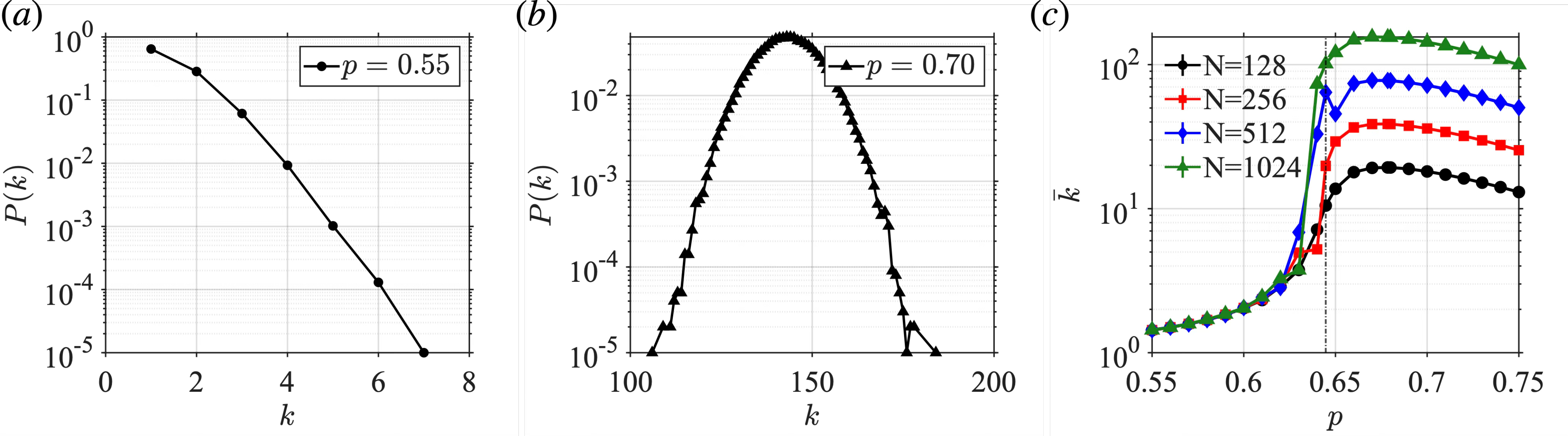}
\caption{Cluster-count statistics of the QSD ($N=1024$ for the distributions; $10^5$ samples per point). \textbf{(a)}~Distribution $P(k)$ of the cluster count in the inactive phase ($p=0.55$), sharply peaked at small $k$. \textbf{(b)}~$P(k)$ in the active phase ($p=0.70$), a broad bell centred at $k\sim140$. \textbf{(c)}~Mean cluster count $\bar k$ versus $p$ for $N=128,256,512,1024$: an $N$-independent $O(1)$ plateau in the inactive phase and an $N$-growing branch in the active phase, separated at $p_c$ (dash-dotted line).}
\label{fig:S-cluster-count}
\end{figure}

The local-clustering analysis of the main text characterises the surviving activity through the number $K(X)=\sum_s n_s(X)$ of clusters (maximal runs of consecutive active sites) in a configuration. Figure~\ref{fig:S-cluster-count} reports its full distribution $P(k)$ and mean $\bar k$ from $10^5$ perfect samples per point.
In the inactive phase [panel~(a), $p=0.55$] $P(k)$ is sharply peaked at small $k$ with an approximately geometric tail, and the mean $\bar k$ is $N$-independent and stays well above unity [panel~(c)]: the surviving activity is a flock built from several clusters rather than a single contiguous run. In the active phase [panel~(b), $p=0.70$] $P(k)$ is a broad bell centred at $k\sim140$ whose support grows with $N$, the bulk-gas behaviour. Panel~(c) shows $\bar k(p)$ for $N=128,256,512,1024$: an $N$-independent $O(1)$ plateau runs through the inactive phase ($p<p_c$) and an $N$-growing branch through the active phase, separated at $p_c$ (dash-dotted line). The persistence of $\bar k>1$ throughout the inactive phase, together with the main-text finite-size result $k_{\rm MI}\to1$, is the geometric signature of the single-flock picture: many counted clusters, one shared position.

\section{Explicit shape function $F(\ell/N)$ from the translated localized-droplet decomposition}
\label{sec:F-shape}

\begin{figure}[h]
\centering
\includegraphics[width=0.9\linewidth]{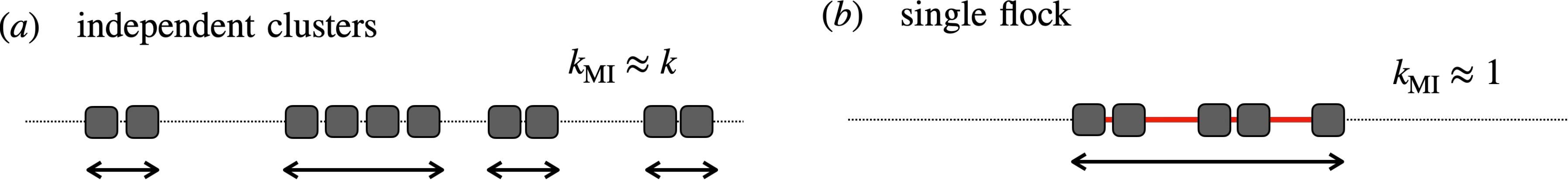}
\caption{Two candidate pictures for the inactive-phase surviving activity. (a) Independent clusters, each free to sit anywhere on the chain: their left--right positions are uncorrelated, so the bipartite information grows with their number, $k_{\rm MI}\approx\bar k$. (b) A single flock: clusters separated by internal holes are strongly correlated into one bound object with a single delocalized position, so the surviving activity encodes only one bit of left--right positional information, $k_{\rm MI}\to 1$. The QSD data identify the inactive phase with~(b); the translated localized-droplet decomposition below formalizes picture~(b).}
\label{fig:S-schematic}
\end{figure}

The QSD density profiles shown in the main text are consistent with an active region forming a single localized cluster of finite transverse extent. This motivates modeling the QSD as a translation-averaged ensemble of a fixed, exponentially localized droplet, from which the bipartite mutual information can be evaluated analytically. The calculation, presented below, yields the asymptotic form
\begin{equation}
I(\ell,N) \;\simeq\; h(\ell/N) \;+\; \frac{K_{\mathrm{eff}}(p)}{N}\,F(\ell/N) \;+\; \frac{1}{N}\,\tilde g(p,N) \;+\; O\!\left(\frac{K_{\mathrm{eff}}^{2}}{N^{2}}\right),
\label{eq:F-shape-form}
\end{equation}
in which the $\ell$-dependent shape is encoded in a closed-form, fully \emph{universal} (i.e., $p$- and $N$-independent) scaling function
\begin{equation}
F(x) \;=\; \log_2\!\bigl[\,x(1-x)\,\bigr],
\label{eq:F-universal}
\end{equation}
all $p$-dependence resides in a single dimensional amplitude $K_{\mathrm{eff}}(p)$, with the interpretation of the droplet's effective half-line width [Eq.~\eqref{eq:K-eff} below], and $\tilde g(p,N)$ is an $\ell$-independent constant eliminated by median subtraction in the bulk. The prediction is verified against numerics at two subcritical points in Sec.~\ref{sec:F-shape-verification}.

\subsection{Assumptions}
\label{sec:F-shape-hypotheses}

Throughout this section we write $\mathbf{x}=(x_1,\ldots,x_N)\in\{0,1\}^N$ for the binary occupation configuration of the chain (the same site variables $x_i$ as in the main text), and reserve $x\equiv\ell/N\in[0,1]$ for the scalar cut fraction. The chain is cut at site $\ell$ into a left subsystem $L_\ell\equiv\{1,\ldots,\ell\}$ of length $\ell$ and a right subsystem $R_\ell\equiv\{\ell+1,\ldots,N\}$ of length $N-\ell$, with $\mathbf{x}_L\equiv(x_1,\ldots,x_\ell)$ and $\mathbf{x}_R\equiv(x_{\ell+1},\ldots,x_N)$ the corresponding configuration restrictions; the bipartite Shannon mutual information $I(\ell,N)\equiv I(\mathbf{x}_L:\mathbf{x}_R)$ studied in the main text is the quantity to be evaluated. As in the main text, all Shannon entropies and mutual informations are measured in bits, i.e., all logarithms are taken base~2. The derivation uses the following structural input on the bond-DP DK quasi-stationary distribution $P_{\mathrm{QS}}$ in the inactive phase.
\begin{enumerate}
\item[(A1)] \emph{Translated localized-droplet decomposition.} There exists a translation-fixed probability measure $\nu$ on $\{0,1\}^{\mathbb{Z}}$ such that
\begin{equation}
P_{\mathrm{QS}}(\mathbf{x}) \;=\; \frac{1}{N}\sum_{z=1}^{N}\nu\!\left(\tau_{-z}\mathbf{x}\right),
\label{eq:F-shape-translated-droplet}
\end{equation}
where $\tau_a$ denotes spatial translation by $a$ sites on the ring. Here $\nu$ is the \emph{canonical droplet distribution}: the conditional law of the QSD given that the flock center sits at the origin. This droplet is the canonical shape of the main-text flock. Concretely, a sample $\tilde{\mathbf{x}}\sim\nu$ specifies the internal active/inactive pattern of one flock anchored at $r=0$; $\nu$ is supported on configurations exponentially localized within transverse extent $\xi_\perp(p)$, and its Shannon entropy $h_\nu\equiv H(\nu)$ measures the internal stochasticity of the droplet (the randomness left after the center position is fixed). The decomposition~\eqref{eq:F-shape-translated-droplet} states that the full QSD is recovered by drawing a center position $z$ uniformly on the ring and laying $\nu$ down translated by $z$. We further take $\nu$ to be generic in the sense that $(z,\tilde{\mathbf{x}})\mapsto\tau_z\tilde{\mathbf{x}}$ is injective, equivalently, $\nu$ has no nontrivial translation stabilizer; this rules out only the measure-zero set of perfectly periodic patterns (e.g.\ $\ldots 1010101 0\ldots$) and is automatic for stochastically generated DK flocks in the inactive phase.
\item[(A2)] \emph{Cut-local center ambiguity.} For distinct center positions $z\ne z'$, the conditional $L$-marginals $\nu(\mathbf{x}_L\mid z)$ and $\nu(\mathbf{x}_L\mid z')$ may overlap on non-vacuum outcomes, but only when at least one of $z,z'$ sits within $O(K_{\mathrm{eff}})$ of the cut $\ell$ (or, equivalently, of the boundary of the droplet support after translation). Verbally: an observer who sees a non-empty droplet imprint deep in the bulk of $L_\ell$ can read off the center position $z$ from the spatial location of the imprint; the only sources of center ambiguity from non-vacuum outcomes are \emph{cut-local} configurations in which the droplet's internal-hole pattern straddles the cut and could equally well be attributed to multiple anchor positions in the cut's $K_{\mathrm{eff}}$-neighborhood. As a consequence, the conditional center entropy decomposes as
\begin{equation}
H(z\mid\mathbf{x}_L) \;=\; P(\mathbf{x}_L=0^\ell)\,\log_2\!\bigl[N\,P(\mathbf{x}_L=0^\ell)\bigr] \;+\; \frac{c_L(p)}{N} \;+\; o(1/N),
\label{eq:F-HZL-decomp}
\end{equation}
where the vacuum branch carries the entire $\ell$-dependent leading $1/N$ correction and $c_L(p)$, capturing the cut-local non-vacuum ambiguities, is bulk-$\ell$-independent by translation symmetry and is absorbed into $\tilde g(p,N)$. The universal shape $F(x)\propto\log_2[x(1-x)]$ derived below therefore originates from the geometry $P(\mathbf{x}_L=0^\ell)\simeq 1-x$ alone, not from any uniqueness-of-decoding requirement on $\nu$; internal collisions modify only the amplitude $K_{\mathrm{eff}}(p)$, the constant $\tilde g$, and the higher-order remainder.
\item[(A3)] \emph{Diluted regime.} $\xi_\perp(p)\ll N$, so an $O(\xi_\perp/N)$ Taylor expansion in the small parameter $K_{\mathrm{eff}}/N$ is controlled (with $K_{\mathrm{eff}}$ defined in~\eqref{eq:K-eff} below).
\end{enumerate}
Assumption (A1) is the rigorous content of standard Yaglom-limit and quasi-stationary existence theorems for one-dimensional subcritical absorbing-state Markov chains~\cite{Ferrari1995,Ferrari1996,Andjel2015,Champagnat2016}; (A2) is generic for droplets carrying internal spatial structure; (A3) is verified directly for the parameters used in the main text ($\xi_\perp\lesssim 6$, $N\ge 128$).

\subsection{Latent-variable entropy identity}
\label{sec:F-shape-identity}

Let $S_L(z),S_R(z)$ denote the conditional Shannon entropies of $\mathbf{x}_L$ and $\mathbf{x}_R$ at fixed center $z$. The entropy chain rule applied to the joint distribution of $(z,\mathbf{x}_L)$ gives
\begin{equation}
H(\mathbf{x}_L) \;=\; \log_2 N \,+\, \langle S_L(z)\rangle_z \,-\, H(z\mid\mathbf{x}_L),
\label{eq:F-HL-chain}
\end{equation}
and similarly for $H(\mathbf{x}_R)$. By (A1), the map $(z,\tilde{\mathbf{x}})\to\mathbf{x}$ is injective, so $H(\mathbf{x})=\log_2 N + h_\nu$ with $h_\nu\equiv H(\nu)$. Substituting into the mutual-information definition $I=H(\mathbf{x}_L)+H(\mathbf{x}_R)-H(\mathbf{x})$,
\begin{equation}
I(\ell,N) \;=\; \log_2 N \;+\; \widehat I_{\mathrm{drop}} \;-\; H(z\mid\mathbf{x}_L) \;-\; H(z\mid\mathbf{x}_R),
\label{eq:F-I-master}
\end{equation}
where
\begin{equation}
\widehat I_{\mathrm{drop}} \;\equiv\; \langle S_L(z)+S_R(z)\rangle_z \,-\, h_\nu \;=\; \frac{1}{N}\sum_{z\in\mathrm{Str}}\!\bigl[S_L(z)+S_R(z)-h_\nu\bigr]
\label{eq:F-Idrop-def}
\end{equation}
is the average $L$--$R$ droplet co-information, restricted to the $O(\xi_\perp)$ ``straddling'' center positions $\mathrm{Str}$ within $\xi_\perp$ of either cut: for non-straddling $z$ one side carries vacuum and $S_L(z)+S_R(z)=h_\nu$ exactly. By rotational symmetry of the ring, $\widehat I_{\mathrm{drop}}$ depends only on $\nu$ near a single cut, not on $\ell$, and is itself of order $\xi_\perp/N$.

\subsection{Vacuum-conditioned center entropy}
\label{sec:F-shape-vac}

The dominant outcome of $\mathbf{x}_L$ on which many $z$ values collide is the vacuum $\mathbf{x}_L=0^{\ell}$, populated by all deep-right center positions plus the $\nu$-fraction of straddling centers whose droplet support fails to reach $L_\ell$. Writing $x\equiv\ell/N$ and
\begin{equation}
K_{\mathrm{eff}}(p) \;\equiv\; \sum_{r=0}^{\infty}\Pr_{\nu}\!\bigl[\,\tilde{\mathbf{x}}\;\text{has nonzero support on}\;\{r+1,r+2,\ldots\}\,\bigr],
\label{eq:K-eff}
\end{equation}
which we interpret as the effective half-line droplet width, the leading-order expansion of $P(\mathbf{x}_L=0^{\ell})$ in $K_{\mathrm{eff}}/N$ is
\begin{equation}
P\bigl(\mathbf{x}_L=0^{\ell}\bigr) \;=\; (1-x) \;-\; \frac{K_{\mathrm{eff}}}{N} \;+\; O\!\Bigl(\!\bigl(K_{\mathrm{eff}}/N\bigr)^{2}\Bigr).
\label{eq:F-Pvac}
\end{equation}
By (A2), the conditional center entropy decomposes into its vacuum branch plus a cut-local, $\ell$-independent piece:
\begin{equation}
H(z\mid\mathbf{x}_L) \;=\; P\bigl(\mathbf{x}_L=0^{\ell}\bigr)\,\log_2\!\Bigl[N\,P\bigl(\mathbf{x}_L=0^{\ell}\bigr)\Bigr] \;+\; \frac{c_L(p)}{N} \;+\; o(1/N).
\label{eq:F-HZL-vac}
\end{equation}
Here $c_L(p)/N$ collects the non-vacuum collision contributions from the $O(K_{\mathrm{eff}})$ straddling region near the cut; by rotational symmetry of the ring, $c_L(p)$ depends only on the droplet shape $\nu$ near a single cut and not on the cut location $\ell$ (in the bulk regime $K_{\mathrm{eff}}\ll\ell\ll N-K_{\mathrm{eff}}$). The symmetric identity holds for $H(z\mid\mathbf{x}_R)$ with $x\leftrightarrow 1-x$ and the corresponding bulk-$\ell$-independent constant $c_R(p)$.

\subsection{First-order Taylor expansion and the shape function $F(x)$}
\label{sec:F-shape-result}

Substituting~\eqref{eq:F-Pvac} into~\eqref{eq:F-HZL-vac} and expanding to first order in $K_{\mathrm{eff}}/N$,
\begin{align}
H(z\mid\mathbf{x}_L) \;&=\; \left[(1{-}x) - \tfrac{K_{\mathrm{eff}}}{N}\right]\!\left[\log_2 N + \log_2(1{-}x) - \tfrac{K_{\mathrm{eff}}}{(1{-}x)\,N}\right] + \frac{c_L(p)}{N} + o(1/N) \nonumber\\
&=\; (1{-}x)\log_2 N \,+\, (1{-}x)\log_2(1{-}x) \,-\, \frac{K_{\mathrm{eff}}}{N}\!\left[\log_2 N \,+\, \log_2(1{-}x) \,+\, 1\right] \nonumber\\
&\qquad \,+\, \frac{c_L(p)}{N} \,+\, o(1/N).
\label{eq:F-HZL-expansion}
\end{align}
The corresponding expansion of $H(z\mid\mathbf{x}_R)$ follows from $x\leftrightarrow 1-x$ with the analogous cut-local constant $c_R(p)$. Summing,
\begin{equation}
\begin{split}
H(z\mid\mathbf{x}_L) + H(z\mid\mathbf{x}_R) \;=\;& \log_2 N \;-\; h(x) \;-\; \frac{K_{\mathrm{eff}}}{N}\!\left[2\log_2 N + \log_2\!\bigl(x(1{-}x)\bigr) + 2\right] \\
&+ \frac{c_L(p)+c_R(p)}{N} + o(1/N).
\end{split}
\label{eq:F-HZsum}
\end{equation}
Inserting~\eqref{eq:F-HZsum} into the latent-variable identity~\eqref{eq:F-I-master} yields the central result of this section,
\begin{equation}
\boxed{\;\; I(\ell,N) \,-\, h(\ell/N) \;=\; \frac{K_{\mathrm{eff}}(p)}{N}\,\log_2\!\bigl[\,x(1{-}x)\,\bigr] \;+\; \frac{1}{N}\,\tilde g(p,N) \;+\; o(1/N), \;\;}
\label{eq:F-shape-main}
\end{equation}
where
\begin{equation}
\tilde g(p,N) \;=\; K_{\mathrm{eff}}(p)\!\left[\,2\log_2 N + 2\,\right] \;+\; N\,\widehat I_{\mathrm{drop}} \;-\; \bigl[c_L(p)+c_R(p)\bigr]
\label{eq:g-tilde-def}
\end{equation}
collects every $\ell$-independent term from~\eqref{eq:F-HZsum}, from the cut-local non-vacuum ambiguities $c_L,c_R$ of~(A2), and from~\eqref{eq:F-Idrop-def}. The $\ell$-dependent shape factor in~\eqref{eq:F-shape-main} is the universal scaling function announced in Eq.~\eqref{eq:F-universal},
\begin{equation}
\boxed{\;\; F(x) \;=\; \log_2\!\bigl[\,x(1-x)\,\bigr]. \;\;}
\label{eq:F-final}
\end{equation}
The $p$- and $N$-independence of $F(x)$ has a transparent origin: $F$ comes \emph{entirely from the vacuum branch} of the conditional center entropy, $P(\mathbf{x}_L=0^\ell)\log_2[N\,P(\mathbf{x}_L=0^\ell)]$ in~\eqref{eq:F-HZL-vac} (and its $L\leftrightarrow R$ counterpart), with both arguments $\simeq 1-x$ and $\simeq x$ at leading order. The shape $\log_2[x(1-x)]$ is therefore a consequence of the macroscopic geometry of the uniform latent center distribution alone, not of any uniqueness-of-decoding statement about non-vacuum droplet imprints. The internal stochasticity of $\nu$ --- including internal holes that produce non-vacuum center ambiguities at the cut --- enters at leading $1/N$ through the cut-local constants $c_L(p),c_R(p)$, which are bulk-$\ell$-independent and thus contribute only to $\tilde g(p,N)$ (and to the subleading remainder), never to the $\ell$-dependent shape $F(x)$. They are eliminated from the shape comparison by median subtraction over the bulk window $0.2\le\ell/N\le 0.8$.

\subsection{Properties of $F(x)$ and the Stirling analogy}
\label{sec:F-shape-properties}

The universal shape function~\eqref{eq:F-final} has three immediate properties:
\begin{itemize}
\item \emph{Left--right symmetry:} $F(x)=F(1-x)$, reflecting the symmetry of the ring under $\ell\leftrightarrow N-\ell$.
\item \emph{Bulk extremum at $x=\tfrac12$:} $F(\tfrac12)=-\log_2 4$ is the maximum value; the overall additive constant is irrelevant since the comparison with data is performed after median subtraction over the bulk window.
\item \emph{Endpoint divergence:} $F(x)\to-\infty$ as $x\to 0$ or $x\to 1$, with leading rate $\log_2 x$. The divergence is regularized at the lattice cutoff $x_{\min}=1/N$, below which the Taylor expansion~\eqref{eq:F-HZL-expansion} ceases to be controlled.
\end{itemize}

The functional form $\log_2[x(1-x)]$ coincides with the standard Stirling correction to the binomial entropy:
\begin{equation}
\frac{1}{N}\log_2\binom{N}{\ell} \;=\; h(\ell/N) \;-\; \frac{1}{2N}\log_2\!\bigl[2\pi N\,x(1-x)\bigr] \;+\; O(N^{-2}),
\label{eq:Stirling-binomial}
\end{equation}
whose $\ell$-dependent part is $-\log_2[x(1-x)]/(2N)$. The localized-droplet correction~\eqref{eq:F-shape-main} reproduces the identical universal shape $\log_2[x(1-x)]$, with the dimensionless prefactor $1/2$ of the binomial-coefficient case replaced by the dimensionful droplet width $K_{\mathrm{eff}}(p)$ characteristic of the QSD. The translated localized-droplet picture thus acts as a discrete-to-continuous correction to the latent-center entropy, with $K_{\mathrm{eff}}$ setting the only $p$-dependent length scale that enters the $O(1/N)$ correction.

\subsection{Numerical verification}
\label{sec:F-shape-verification}

Equation~\eqref{eq:F-shape-main} predicts that, after subtracting the leading $h(\ell/N)$ and the $\ell$-independent constant $\tilde g(p,N)$, the rescaled deviation $N[I(\ell,N)-h(\ell/N)] - \tilde g(p,N)$ collapses across both $N$ and $p$ onto the single universal curve $K_{\mathrm{eff}}(p)\,F(x)=K_{\mathrm{eff}}(p)\,\log_2[x(1-x)]$. We verify this at two subcritical points, $p=0.55$ and $p=0.60$, with chain lengths $N\in\{128,256,512,1024\}$.

\begin{figure}[h]
\centering
\includegraphics[width=0.7\linewidth]{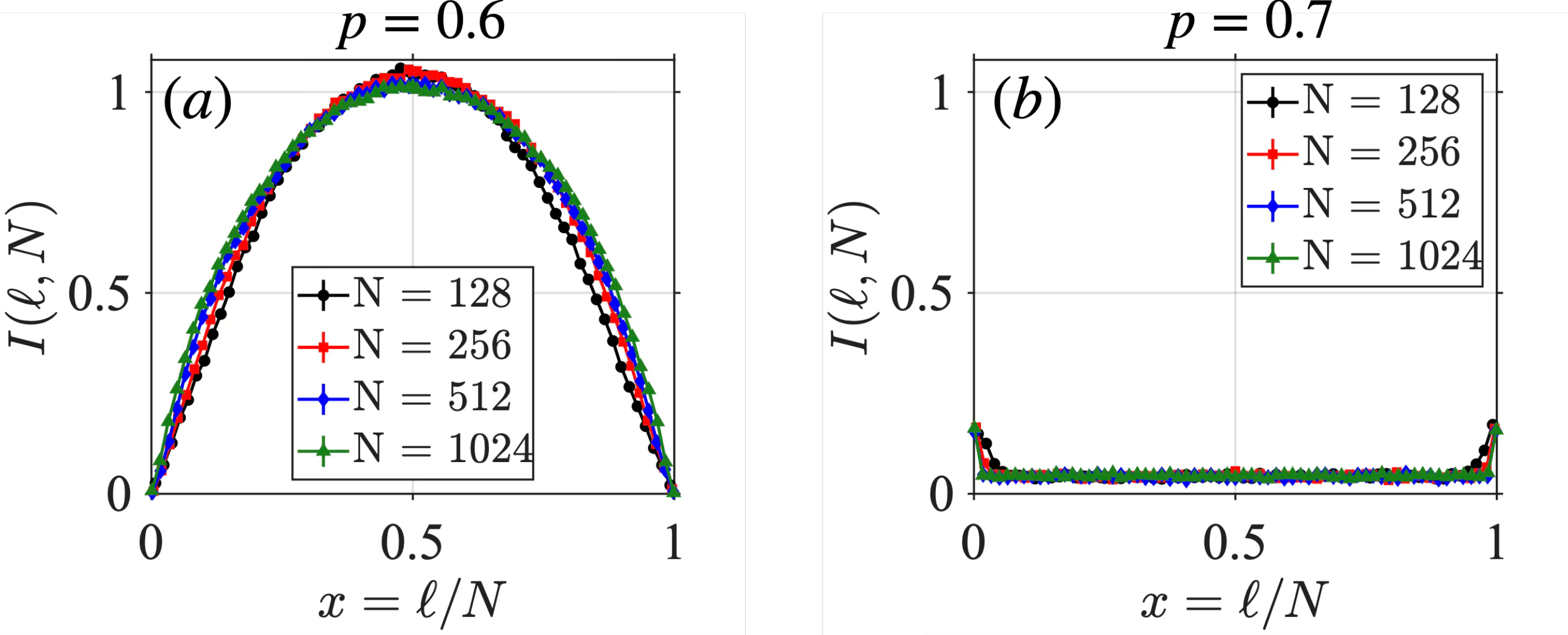}
\caption{Bipartite mutual information $I(\ell,N)$ of the QSD (in bits, $\log_2$) versus cut fraction $x=\ell/N$ for $N=128,256,512,1024$. \textbf{(a)}~Inactive phase ($p=0.60$): the curves approach the single-binary-trial entropy $h(x)$ from above as $N$ grows, peaking at $h(1/2)=1$; the residual $I-h\sim K_{\mathrm{eff}}/N$ is analysed in this section. \textbf{(b)}~Active phase ($p=0.70$): $I(\ell,N)$ is essentially zero across all cuts, signalling the absence of an extended positional degree of freedom.}
\label{fig:S-mi-raw}
\end{figure}

Before the rescaled collapse, Fig.~\ref{fig:S-mi-raw} shows the raw bipartite mutual information underlying the analysis: in the inactive phase [panel~(a), $p=0.60$] the bond-resolved $I(\ell,N)$ approaches the single-bit form $h(\ell/N)$ from above as $N$ grows and peaks at $h(1/2)=1$, while in the active phase [panel~(b), $p=0.70$] it is essentially zero across all cuts --- the full-cut counterpart of the half-chain inset in the main-text mutual-information figure.

For each $(p,N)$ we extract the amplitude $K_{\mathrm{eff}}$ from a least-squares fit of the median-subtracted $N[I-h]-\tilde g$ against $F(x)=\log_2[x(1-x)]$ on the fit window $0.05\le x\le 0.95$, with the basis centered on the same window to make the slope $K_{\mathrm{eff}}$ invariant under additive shifts of $F$. The pooled fits give
\begin{align}
p=0.55:&\qquad K_{\mathrm{eff}}\;=\;12.4\;\text{sites},\qquad \mathrm{RMS}_{\mathrm{resid}}\;=\;3.6, \nonumber\\
p=0.60:&\qquad K_{\mathrm{eff}}\;=\;15.1\;\text{sites},\qquad \mathrm{RMS}_{\mathrm{resid}}\;=\;6.0,
\label{eq:K-eff-numerical}
\end{align}
with per-$N$ fits $K_{\mathrm{eff}}^{(N)}\in\{11.3,\,\allowbreak 12.3,\,\allowbreak 13.9,\,\allowbreak 12.3\}$ at $p=0.55$ (spread $\pm 11\%$) and $\{13.0,\,\allowbreak 19.4,\,\allowbreak 18.7,\,\allowbreak 7.9\}$ at $p=0.60$ (the $N=1024$ point at $p=0.60$ is dominated by sampling noise since the absolute signal $I-h\sim K_{\mathrm{eff}}/N\sim 10^{-2}$ is comparable to the per-cut standard error). The two $K_{\mathrm{eff}}$ values track the $p$-dependence of the underlying droplet width and grow as $p\to p_c$, but the $\ell$-dependent shape $F(x)$ is the same universal $\log_2[x(1-x)]$ at both points.

\begin{figure}[h]
\centering
\includegraphics[width=0.47\linewidth]{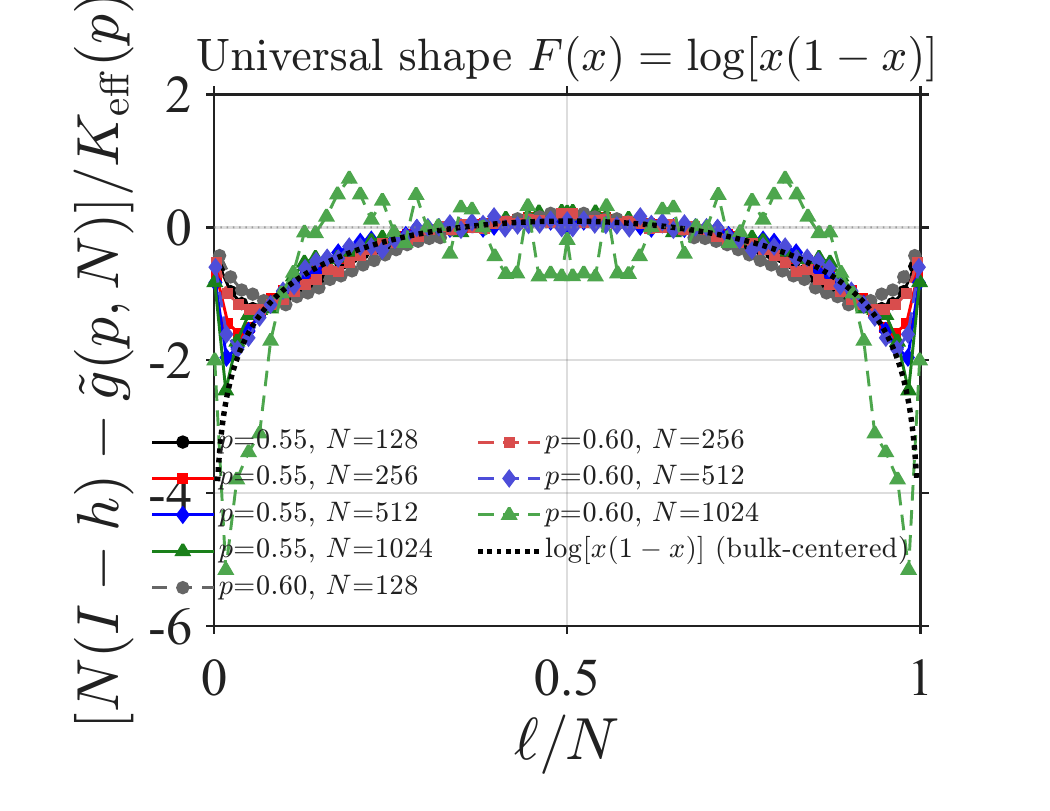}\hfill
\includegraphics[width=0.50\linewidth]{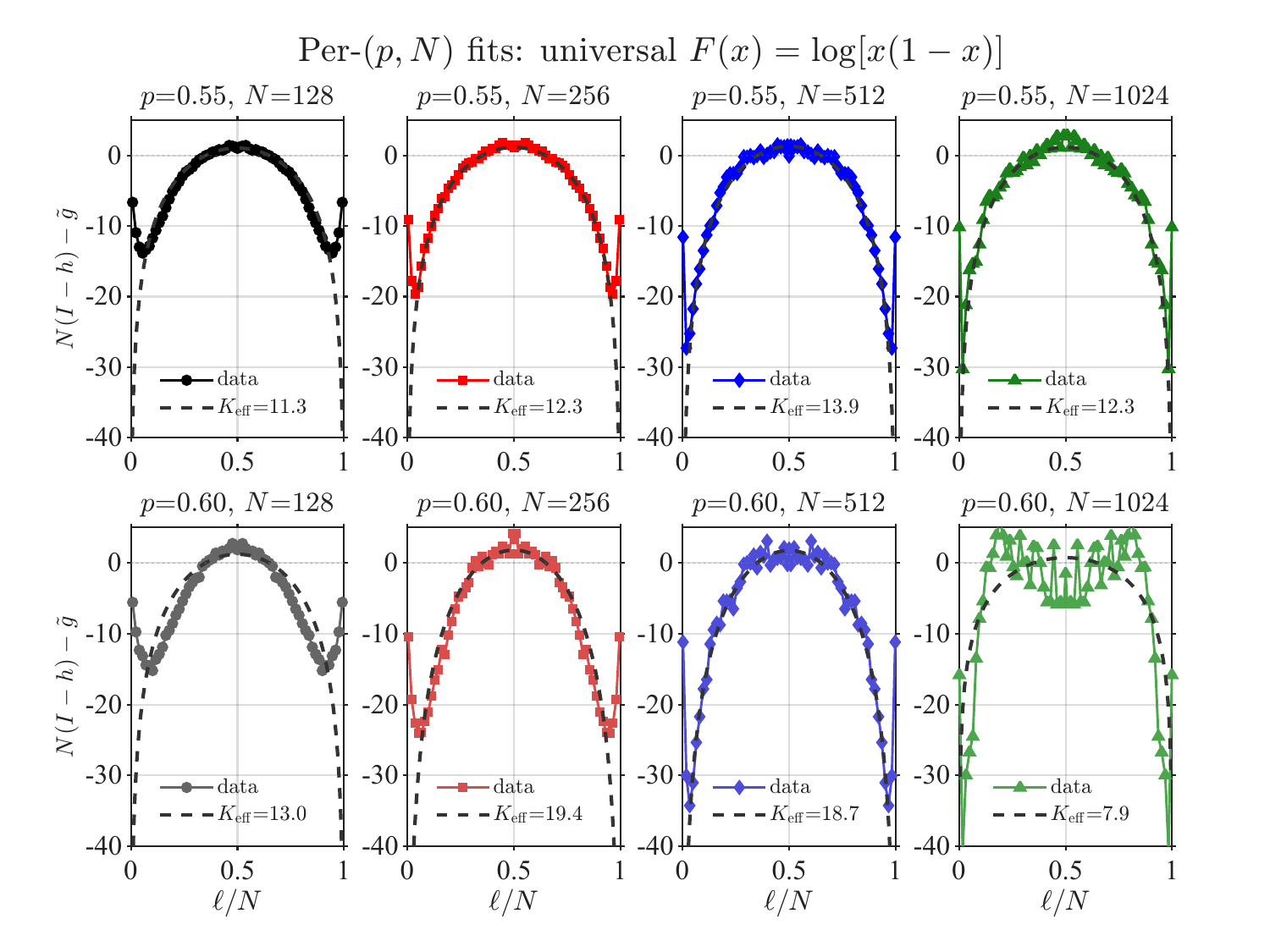}
\caption{Universality of the analytic shape $F(x)=\log_2[x(1{-}x)]$. \textbf{Left:} median-subtracted residual divided by the fitted amplitude, $\{N[I-h]-\tilde g(p,N)\}/K_{\mathrm{eff}}(p)$, for $p\in\{0.55,\,0.60\}$ and $N\in\{128,256,512,1024\}$; eight curves overlap onto the single universal $\log_2[x(1-x)]$ (dashed). \textbf{Right:} per-$(p,N)$ panels showing the extracted $K_{\mathrm{eff}}^{(N)}(p)$ for each chain length; the $\ell$-dependent shape is universal while the amplitude is $p$-dependent.}
\label{fig:S-F-ansatz}
\end{figure}

The universal collapse is visualised in Fig.~\ref{fig:S-F-ansatz}: the left panel overlays the median-subtracted shape divided by $K_{\mathrm{eff}}(p)$ for both $p$ values onto the single universal curve $\log_2[x(1-x)]$ (dashed); the right panel gives per-$(p,N)$ residual plots with the extracted $K_{\mathrm{eff}}^{(N)}$ printed in each subpanel. Bulk and endpoint regions of every curve track the universal $\log_2[x(1-x)]$ to within sampling noise.

\subsection{Rigor, range of validity, and limitations}
\label{sec:F-shape-rigor}

The derivation of~\eqref{eq:F-final} from (A1)--(A3) consists of the following steps:
\begin{enumerate}
\item The latent-variable identity~\eqref{eq:F-I-master} is the exact entropy chain rule applied to (A1); no approximation enters.
\item The decomposition of $H(z\mid\mathbf{x}_L)$ in~\eqref{eq:F-HZL-vac} separates the vacuum branch (which carries the entire $\ell$-dependent $1/N$ contribution) from a bulk-$\ell$-independent cut-local piece $c_L(p)/N$ supplied by (A2).
\item The leading-order form of $P(\mathbf{x}_L=0^{\ell})$ in~\eqref{eq:F-Pvac} follows from counting deep-right and straddling-zero center positions; the error is $O((K_{\mathrm{eff}}/N)^{2})$.
\item The Taylor expansion~\eqref{eq:F-HZL-expansion} is Lagrange-controlled in (A3).
\item Translation invariance of the ring guarantees that $\widehat I_{\mathrm{drop}}$ and the cut-local constants $c_L(p),c_R(p)$ are all $\ell$-independent and contribute only to $\tilde g(p,N)$.
\end{enumerate}
Each step is rigorous at leading order in $K_{\mathrm{eff}}/N$, and the universal shape~\eqref{eq:F-final} follows under (A1)--(A3) as a controlled $O(K_{\mathrm{eff}}/N)$ identity for $C\,K_{\mathrm{eff}}\le\ell\le N-C\,K_{\mathrm{eff}}$ with $C$ a hypothesis-dependent constant. Crucially, no statement about uniqueness of decoding of non-vacuum droplet imprints is required: the universal log-shape of $F(x)$ originates from the geometry of the vacuum branch alone, while droplet internal collisions modify only the bulk-$\ell$-independent constants.

Three caveats remain:
\begin{enumerate}
\item \emph{Lattice cutoff.} The Taylor expansion~\eqref{eq:F-HZL-expansion} requires $\ell\gtrsim K_{\mathrm{eff}}$. For $\ell\lesssim K_{\mathrm{eff}}$ the cut sits inside the droplet and the leading-order picture breaks down; the data deviates from the ansatz at $\ell=1$ visible in Fig.~\ref{fig:S-F-ansatz}.
\item \emph{Definition of $K_{\mathrm{eff}}$.} Equation~\eqref{eq:K-eff} defines $K_{\mathrm{eff}}(p)$ intrinsically from the droplet distribution $\nu$; extracting it independently from the connected correlator $G(r)$ would require model-specific input on $\nu$ that we do not pursue here. The values reported in~\eqref{eq:K-eff-numerical} are therefore the direct empirical determinations of $K_{\mathrm{eff}}(p)$ from the bipartite-MI data itself. The data confirm that $K_{\mathrm{eff}}(p)$ and the connected-correlator length $\xi_\perp(p)$ are distinct length scales: their ratio $K_{\mathrm{eff}}/\xi_\perp$ takes different values at $p=0.55$ ($\simeq 4.8$, with $\xi_\perp=2.59$) and $p=0.60$ ($\simeq 2.6$, with $\xi_\perp=5.9$), so the droplet width is \emph{not} simply proportional to the connected-correlator length.
\item \emph{Predicted $\tilde g(p,N)$.} Equation~\eqref{eq:F-shape-main} predicts a leading $\tilde g(p,N)/N\sim 2K_{\mathrm{eff}}\log_2 N/N$ growth, whereas the empirically extracted $\tilde g$ values do not display the expected logarithm with $N$. This indicates that the deep/straddle boundary~\eqref{eq:F-Idrop-def} carries additional $O(K_{\mathrm{eff}}/N)\log_2 N$ terms that partially cancel the explicit logarithm in~\eqref{eq:F-shape-main}; since they are $\ell$-independent, they affect only $\tilde g(p,N)$ and not the universal shape $F(x)$.
\end{enumerate}
The universal shape result~\eqref{eq:F-final} is robust to these caveats: it follows from the universal structure of $P(\mathbf{x}_L=0^{\ell})\sim 1-x$ and $P(\mathbf{x}_R=0^{N-\ell})\sim x$, both of which are direct consequences of the uniform latent-center distribution implied by the translated-droplet hypothesis (A1); no microscopic detail of $\nu$ enters the $\ell$-dependent part of the $O(1/N)$ correction.

\end{appendix}

\bibliography{references}

\begin{thebibliography}{10}
\providecommand{\url}[1]{\texttt{#1}}
\providecommand{\urlprefix}{URL }
\expandafter\ifx\csname urlstyle\endcsname\relax
  \providecommand{\doi}[1]{doi:\discretionary{}{}{}#1}\else
  \providecommand{\doi}{doi:\discretionary{}{}{}\begingroup
  \urlstyle{rm}\Url}\fi
\providecommand{\eprint}[2][]{\url{#2}}

\bibitem{Janssen1981}
H.~K. Janssen,
\newblock \emph{On the nonequilibrium phase transition in reaction-diffusion
  systems with an absorbing stationary state},
\newblock Zeitschrift f{\"u}r Physik B Condensed Matter \textbf{42}(2), 151
  (1981),
\newblock \doi{10.1007/BF01319549}.

\bibitem{Grassberger1982}
P.~Grassberger,
\newblock \emph{On phase transitions in schl{\"o}gl's second model},
\newblock Zeitschrift f{\"u}r Physik B Condensed Matter \textbf{47}(4), 365
  (1982),
\newblock \doi{10.1007/BF01313803}.

\bibitem{Hinrichsen2000}
H.~Hinrichsen,
\newblock \emph{Non-equilibrium critical phenomena and phase transitions into
  absorbing states},
\newblock Advances in Physics \textbf{49}(7), 815 (2000),
\newblock \doi{10.1080/00018730050198152}.

\bibitem{Domany1984}
E.~Domany and W.~Kinzel,
\newblock \emph{Equivalence of cellular automata to {Ising} models and directed
  percolation},
\newblock Physical Review Letters \textbf{53}(4), 311 (1984),
\newblock \doi{10.1103/PhysRevLett.53.311}.

\bibitem{Dickman2002}
R.~Dickman and R.~Vidigal,
\newblock \emph{Quasi-stationary distributions for stochastic processes with an
  absorbing state},
\newblock Journal of Physics A: Mathematical and General \textbf{35}(5), 1147
  (2002),
\newblock \doi{10.1088/0305-4470/35/5/303}.

\bibitem{Oliveira2005}
M.~M. de~Oliveira and R.~Dickman,
\newblock \emph{How to simulate the quasistationary state},
\newblock Physical Review E \textbf{71}(1), 016129 (2005),
\newblock \doi{10.1103/PhysRevE.71.016129}.

\bibitem{Harada2019}
K.~Harada and N.~Kawashima,
\newblock \emph{Entropy governed by the absorbing state of directed
  percolation},
\newblock Physical Review Letters \textbf{123}(9), 090601 (2019),
\newblock \doi{10.1103/PhysRevLett.123.090601}.

\bibitem{Pizzi2022}
A.~Pizzi, D.~Malz, A.~Nunnenkamp and J.~Knolle,
\newblock \emph{Bridging the gap between classical and quantum many-body
  information dynamics},
\newblock Physical Review B \textbf{106}(21), 214303 (2022),
\newblock \doi{10.1103/PhysRevB.106.214303}.

\bibitem{Pizzi2024}
A.~Pizzi and N.~Y. Yao,
\newblock \emph{Bipartite mutual information in classical many-body dynamics},
\newblock Physical Review B \textbf{110}(2), L020301 (2024),
\newblock \doi{10.1103/PhysRevB.110.L020301}.

\bibitem{Jensen2004}
I.~Jensen,
\newblock \emph{Low-density series expansions for directed percolation: {III}.
  some two-dimensional lattices},
\newblock Journal of Physics A: Mathematical and General \textbf{37}(27), 6899
  (2004),
\newblock \doi{10.1088/0305-4470/37/27/003}.

\bibitem{ChetriteTouchette2015}
R.~Chetrite and H.~Touchette,
\newblock \emph{Nonequilibrium {M}arkov processes conditioned on large
  deviations},
\newblock Annales Henri Poincar{\'{e}} \textbf{16}(9), 2005 (2015),
\newblock \doi{10.1007/s00023-014-0375-8}.

\bibitem{Collet2013}
P.~Collet, S.~Mart{\'{i}}nez and J.~San~Mart{\'{i}}n,
\newblock \emph{Quasi-Stationary Distributions: {M}arkov Chains, Diffusions and
  Dynamical Systems},
\newblock Probability and Its Applications. Springer, Berlin, Heidelberg,
\newblock \doi{10.1007/978-3-642-33131-2} (2013).

\bibitem{Harada2025}
K.~Harada, T.~Okubo and N.~Kawashima,
\newblock \emph{Tensor tree learns hidden relational structures in data to
  construct generative models},
\newblock Machine Learning: Science and Technology \textbf{6}(2), 025002
  (2025),
\newblock \doi{10.1088/2632-2153/adc2c7}.

\bibitem{Ferrari1995}
P.~A. Ferrari, H.~Kesten, S.~Mart{\'{i}}nez and P.~Picco,
\newblock \emph{Existence of quasi-stationary distributions. {A} renewal
  dynamical approach},
\newblock The Annals of Probability \textbf{23}(2), 501 (1995),
\newblock \doi{10.1214/aop/1176988277}.

\bibitem{Ferrari1996}
P.~A. Ferrari, H.~Kesten and S.~Mart{\'{i}}nez,
\newblock \emph{$r$-positivity, quasi-stationary distributions and ratio limit
  theorems for a class of probabilistic automata},
\newblock The Annals of Applied Probability \textbf{6}(2), 577 (1996),
\newblock \doi{10.1214/aoap/1034968146}.

\bibitem{Champagnat2016}
N.~Champagnat and D.~Villemonais,
\newblock \emph{Exponential convergence to quasi-stationary distribution and
  {$Q$}-process},
\newblock Probability Theory and Related Fields \textbf{164}(1--2), 243 (2016),
\newblock \doi{10.1007/s00440-014-0611-7}.

\bibitem{Derrida1993}
B.~Derrida, M.~R. Evans, V.~Hakim and V.~Pasquier,
\newblock \emph{Exact solution of a {1D} asymmetric exclusion model using a
  matrix formulation},
\newblock Journal of Physics A: Mathematical and General \textbf{26}(7), 1493
  (1993),
\newblock \doi{10.1088/0305-4470/26/7/011}.

\bibitem{Johnson2010}
T.~H. Johnson, S.~R. Clark and D.~Jaksch,
\newblock \emph{Dynamical simulations of classical stochastic systems using
  matrix product states},
\newblock Physical Review E \textbf{82}(3), 036702 (2010),
\newblock \doi{10.1103/PhysRevE.82.036702}.

\bibitem{Banuls2019}
M.~C. Ba{\~{n}}uls and J.~P. Garrahan,
\newblock \emph{Using matrix product states to study the dynamical large
  deviations of kinetically constrained models},
\newblock Physical Review Letters \textbf{123}(20), 200601 (2019),
\newblock \doi{10.1103/PhysRevLett.123.200601}.

\bibitem{Helms2019}
P.~Helms, U.~Ray and G.~K.-L. Chan,
\newblock \emph{Dynamical phase behavior of the single- and multi-lane
  asymmetric simple exclusion process via matrix product states},
\newblock Physical Review E \textbf{100}(2), 022101 (2019),
\newblock \doi{10.1103/PhysRevE.100.022101}.

\bibitem{Boesl2026}
J.~Boesl, F.~Pollmann and M.~Knap,
\newblock \emph{Entanglement pattern transition of quantum states from directed
  percolation},
\newblock arXiv preprint arXiv:2605.26219  (2026),
\newblock \eprint{2605.26219}.

\bibitem{Andjel2015}
E.~Andjel, F.~Ezanno, P.~Groisman and L.~T. Rolla,
\newblock \emph{Subcritical contact process seen from the edge: convergence to
  quasi-equilibrium},
\newblock Electronic Journal of Probability \textbf{20}(32), 1 (2015),
\newblock \doi{10.1214/EJP.v20-3881}.

\end{thebibliography}

\end{document}